\documentclass[final]{jfm}
\usepackage{graphicx}
\usepackage{cancel}

\usepackage{epstopdf}

\usepackage{amsmath,amssymb,amsfonts}
\usepackage[mathscr]{eucal}
\usepackage{comment}
\usepackage{multirow}

\usepackage{soul}
\usepackage{cancel}
\usepackage[normalem]{ulem}

\usepackage{subfig}
\usepackage{natbib}

\usepackage{hyperref}
\hypersetup{
	breaklinks,
	colorlinks=true,
	linkcolor=blue,
	citecolor=Purple,
	pdfauthor={N C Constantinou and W R Young},
	pdftitle={Beta-plane turbulence above monoscale topography}
}

\usepackage{microtype}

\usepackage{bm,xspace}

\usepackage[usenames,dvipsnames]{xcolor}

\newcommand{\MATLAB}{\textsc{Matlab}\xspace}

\newcommand{\quasilinear}{$\rm QL$}

\newcommand{\beq}{\begin{equation} }

\newcommand{\eeq}{\end{equation} }

\newcommand{\psit}{\psi^{\mathrm{test}}}
\newcommand{\Umin}{\bar U^{\mathrm{min}}}
\newcommand{\UminE}{\bar U_E^{\mathrm{min}}}
\newcommand{\UminQ}{\bar U_Q^{\mathrm{min}}}

\newcommand{\corr}{\mathop{\rm corr}\nolimits}

\newcommand{\defn}{\ensuremath{\stackrel{\mathrm{def}}{=}}}

\def\beq{\begin{equation}}
\def\eeq{\end{equation}}

\newcommand{\bx}{\boldsymbol{x}}

\newcommand{\bU}{\boldsymbol{U}}

\newcommand{\bk}{\boldsymbol{k}}

\newcommand{\J}{\boldsymbol{\mathsf{J}}}

\newcommand{\bE}{\ensuremath {\boldsymbol {E}}}

\newcommand{\leta}{\ell_\eta}
\newcommand{\Leta}{L_\eta}

\newcommand{\etarms}{\eta_{\mathrm{rms}}}

\newcommand{\mueff}{\mu_{\mathrm{eff}}}
\newcommand{\keff}{\kappa_{\mathrm{eff}}}

\providecommand\bnabla{\boldsymbol{\nabla}}
\providecommand\bcdot{\boldsymbol{\cdot}}

\def\wind{F}

\def\ii{{\rm i}}
\def\dd{{\rm d}}
\def\ee{{\rm e}}

\newcommand{\D}{\mathrm{D}}

\renewcommand{\S}{\mathcal{S}}
\renewcommand{\P}{\mathcal{P}}

\def\la{\langle}
\def\ra{\rangle}

\def\laa{\left\langle}
\def\raa{\right\rangle}

\def\Fs{F/(\mu\etarms\leta)}
\def\bs{\beta\leta/\etarms}
\def\mus{\mu/\etarms}

\def\Fs{F_*}
\def\bs{\beta_*}
\def\mus{\mu_*}

\def\bit{\vphantom{\dot{W}}}
\newcommand{\lap}{{\nabla^2}} 

\newcommand{\half }{\tfrac{1}{2}}

\newcommand{\grad}{\ensuremath{\boldsymbol {\nabla}}}

\renewcommand{\J}{\mathsf{J}}
\renewcommand{\J}{\mathrm{J}} 

\newcommand{\com}{\, ,}
\newcommand{\per}{\, .}
\newcommand{\andd}{\qquad \text{and} \qquad }

\newcommand{\z}{\zeta}
\newcommand{\h}{\eta}
\renewcommand{\(}{\left(}
\renewcommand{\[}{\left[}
\renewcommand{\)}{\right)}
\renewcommand{\]}{\right]}

\def\XXint#1#2#3{{\setbox0=\hbox{$#1{#2#3}{\int}$}
     \vcenter{\hbox{$#2#3$}}\kern-.5\wd0}}

\def\fs{\laa\psi\eta_x\raa}
\def\fsb{\la\bar{\psi}\eta_x\ra}

\shorttitle{Beta-plane turbulence above monoscale topography}
\shortauthor{Constantinou and Young}

\title{Beta-plane turbulence above monoscale topography}

\author{Navid C. Constantinou\aff{1}
  \corresp{\email{navid@ucsd.edu}},
 \and  William R. Young\aff{1}}

\affiliation{\aff{1}Scripps Institution of Oceanography, University of California San Diego, La Jolla, CA~92093-0213, USA}

\begin{document}

\maketitle

\begin{abstract}


	Using a one-layer quasi-geostrophic model, we study the effect of  random monoscale topography on forced beta-plane turbulence. The forcing is  a uniform steady wind stress that produces both  a uniform large-scale zonal flow $U(t)$ and smaller-scale macroturbulence  characterized by both standing and transient eddies. The large-scale flow $U$ is  retarded by a combination of Ekman drag  and the domain-averaged topographic form stress produced by the eddies. The topographic form stress typically balances most of the applied wind stress, while the Ekman drag provides all of the energy dissipation required to balance the wind work. A collection of statistically equilibrated numerical solutions delineates the main flow regimes and the dependence of the time-average of $U$ on parameters such as the planetary vorticity gradient $\beta$  and the statistical properties of the topography. We obtain asymptotic scaling laws for the strength of the large-scale flow $U$ in the limiting cases of  weak and strong forcing.


	If $\beta$ is significantly smaller  than the topographic PV gradient then the flow consists of  stagnant pools attached to pockets of closed geostrophic contours. The stagnant dead zones are bordered by  jets and the flow through the domain  is concentrated into a narrow channel of open geostrophic contours. In most of the domain the flow is weak and thus  the large-scale flow $U$ is an unoccupied mean.

	If $\beta$ is comparable to, or larger than, the topographic PV gradient then all geostrophic contours are open and the flow is uniformly distributed throughout the domain. In this open-contour case there is  an ``eddy saturation'' regime in which  $U$ is insensitive to large changes in the wind stress. We show that eddy saturation requires strong transient eddies that act effectively as PV diffusion. This PV diffusion does not alter the kinetic energy of the standing eddies, but it does increase the topographic form stress by enhancing the correlation between topographic slope and the standing-eddy pressure field. Using bounds based on the energy and enstrophy power integrals we show that as the strength of the wind stress increases the flow transitions from a regime in which most of the form stress balances the wind stress to a regime in which the form stress is very small and  large transport ensues.
\end{abstract}


\vspace{-4em}

\section{Introduction\label{sec:intro}}
Winds force the oceans by applying a stress at the sea surface. A question of interest is where and how this vertical flux of horizontal momentum into the ocean is balanced. Consider, for example, a steady zonal wind  blowing over the sea surface and exerting a force on the ocean. In a statistically steady state we can identify all possible mechanisms for balancing this surface force by first vertically integrating over the depth of the ocean, and then horizontally integrating over a region in which the wind stress is approximately uniform. Following the strategy of~\citet{Bretherton-Karweit-1975}, we have in mind a mid-ocean region which is much smaller than ocean basins, but much larger than the length scale of ocean macroturbulence. The zonal wind stress on  this volume  can be balanced by several processes which we classify as either local or non-local. The most obvious local process is Ekman  drag  in turbulent bottom boundary layers. But in the deep ocean Ekman drag is negligible \citep{Munk-Palmen-1951}; instead the most important local process is topographic form stress (the correlation of pressure and topographic slope). Topographic form stress is an inviscid mechanism for coupling the ocean to the solid Earth. Non-local processes include the advection of zonal momentum out of the domain and, most importantly, the possibility that a large-scale  pressure gradient is supported  by piling water up against either distant continental boundaries or ridge systems.


In this paper we concentrate on the local processes that balance wind stress and result in homogeneous ocean macroturbulence. Thus we investigate the simplest model of topographic form stress. This is a single-layer quasi-geostrophic~(QG) model, forced by a steady zonal wind stress in a doubly periodic domain \citep{Hart-1979,Davey-1980a,Holloway-1987,Carnevale-Frederiksen-1987}. A distinctive feature of the model is a uniform large-scale zonal flow $U(t)$ that is accelerated by the applied uniform surface wind stress $\tau$ while resisted by both Ekman bottom drag $\mu U$ and domain-averaged topographic form stress:
\beq
	U_t = \wind - \mu U  - \la\psi \h_x\ra \per
	\label{eq:U_t}
\eeq
In \eqref{eq:U_t}, $\wind=\tau/(\rho_0 H)$ where  $\rho_0$ is the reference density of the fluid and $H$ is the mean depth. The eddy streamfunction $\psi(x,y,t)$ in \eqref{eq:U_t} evolves according to the quasi-geostrophic potential vorticity (QGPV) equation~\eqref{eq:z_NL}, $\h$ is the topographic PV and $\fs$ is the domain-averaged topographic form stress. (All quantities are fully defined in section~\ref{sec:form}.)

\renewcommand{\arraystretch}{1.35}
\begin{table}
	\vspace{-1em}
\caption{Various idealized topographies previously used in the literature. }
	\vspace{-1em}\begin{center}
\begin{tabular}{l c} \\
	\cite{Charney-DeVore-79} & $\cos{(m\upi x)}\sin{(n\upi y)}$ \\
	\cite{Charney-etal-1981} & $h(x)\sin{(\upi y)}$\\
	\cite{Hart-1979} & $\cos{(2\upi x)}$ (plus some remarks on $h(y)\cos{(2\upi x)}$)\\
	\cite{Davey-1980a} & triangular ridge:~$h(x)\sin{(\upi y)}$\\
	\cite{Pedlosky-1981} & $\cos{(m \upi x)}\sin{(n\upi y)}$\\
	\cite{Kallen-1982} & $P_3^2(r)\cos(3\phi)$ (on the sphere)\\
	\cite{Rambaldi-Flierl-1983}& $\sin{(2\upi x)}$\\
	\cite{Rambaldi-Mo-1984} & $\sin{(\upi y)}\sin{(4\upi x)}$\\
	\cite{Yoden-1985} & $\cos{(m\upi x)}\sin{( n \upi y)}$\\
	\cite{Legras-Ghil-1985} & $P_2^1(r)\cos(2\phi)$ (on the sphere)\\
	\cite{Tung-Rosenthal-1985} & $\cos{(m \upi x)}\sin{(n \upi y)}$ \\
	\cite{Uchimoto-Kubokawa-2005} & $\sin{(2 \upi x)}\sin{(\upi x)}$\\
\end{tabular}
\end{center}\label{tab:prev_lit}
\end{table}

This model may be  pertinent to the Southern Ocean. There, the absence of continental boundaries along a range of latitudes implies that a large-scale pressure gradient cannot be invoked in balancing the zonal wind stress. However, we emphasize that the model in~\eqref{eq:U_t} and~\eqref{eq:z_NL} may also be relevant in a small region of the ocean away from any continental boundaries, where we expect a statistically homogeneous eddy field. Although the model has been derived previously by several authors, it has never been investigated in detail except under severe low-order spectral truncation, and only for the simplest model topographies summarized in table~\ref{tab:prev_lit}. Here, we delineate the various flow regimes of geostrophic turbulence above a homogeneous, isotropic and monoscale topography e.g., the topography shown in figure~\ref{fig:topo}.

Similar models were developed in meteorology in order to understand stationary waves and blocking patterns. \citet{Charney-DeVore-79} introduced a reduced model of the interaction of zonal flow and topography and demonstrated the possibility of multiple equilibrium states, one of which corresponds to a topographically blocked flow. \citet{Charney-DeVore-79} paved the way for the studies summarized in table~\ref{tab:prev_lit}, which are directed at understanding the existence of multiple stable solutions to systems such as~\eqref{eq:U_t}. This meteorological literature is mainly concerned with planetary-scale topography e.g., note the use of low-order spherical harmonics and small wavenumbers in table~\ref{tab:prev_lit}. Here, reflecting our interest in oceanographic issues, we consider smaller scale topography such as features with 10~to~$100\,\textrm{km}$ scale i.e., topography with roughly the same scale as ocean macroturbulence. Despite this difference, we also find a regime with multiple stable states and hysteresis (section~\ref{survey}).

\begin{figure}
\centering
\includegraphics[width = .8\textwidth]{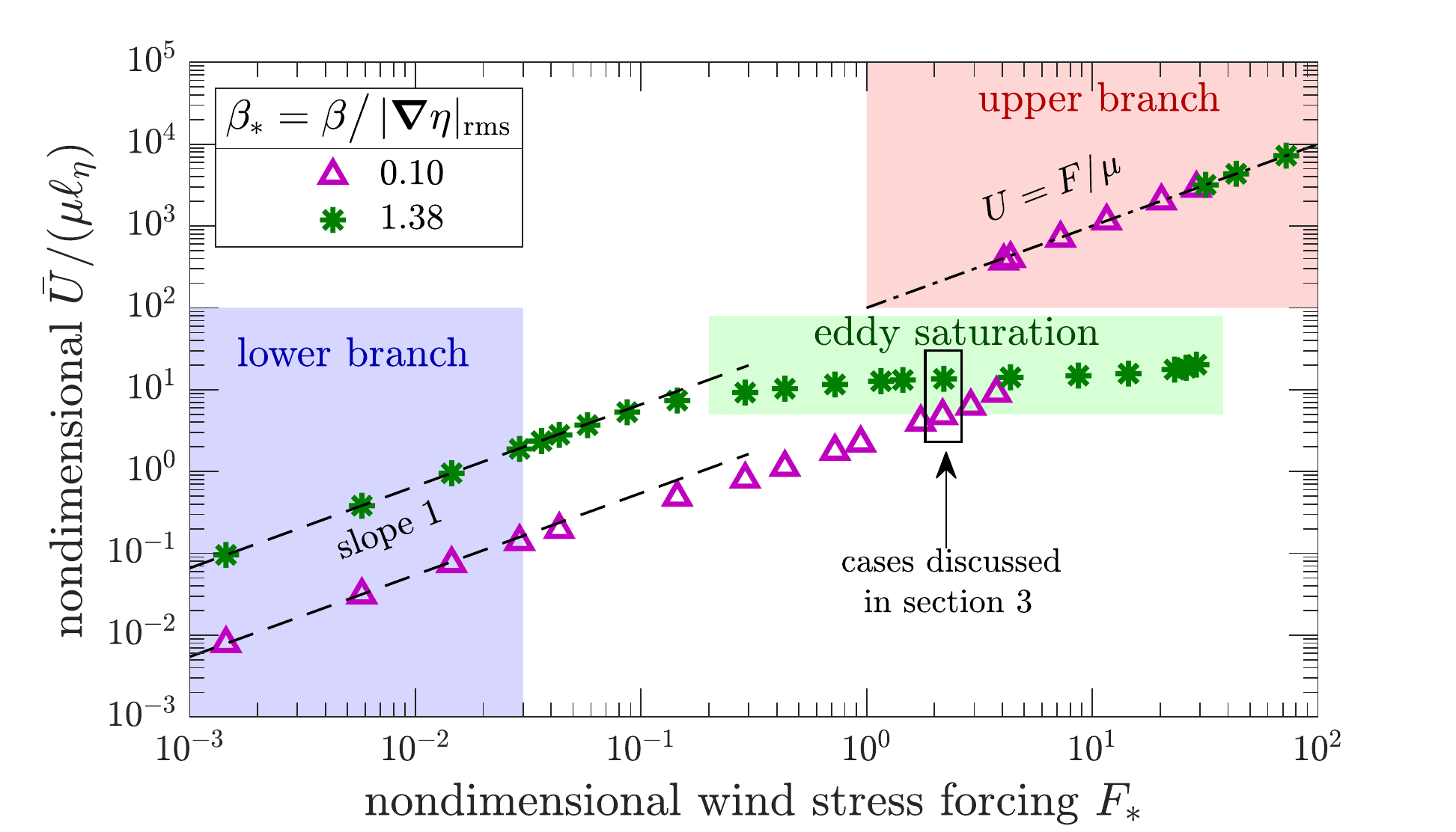}
\caption{Dependence of time-mean large-scale flow $\bar U$ on  wind stress forcing $F$. The parameters $\bs$ and $\Fs$ are defined in section~\ref{sec:nondim}. The box encloses the two points  discusses in sections \ref{sec:example_closed} and \ref{sec:example_open}.}\label{figIntro}
\end{figure}

Figure~\ref{figIntro} summarizes our main result by showing how the time-mean large-scale flow  $\bar U$ varies with increasing wind stress forcing $F$. The two solution suites  shown in figure~\ref{figIntro} represent two end-points corresponding to either closed geostrophic contours (small value of $\bs$, which is the ratio of the planetary PV gradient to the r.m.s.~topographic PV gradient) or open geostrophic contours (large $\bs$). In both cases there are  two flow regimes in which the flow is steady without transient eddies: the ``lower branch'' and the ``upper branch'' (indicated in figure~\ref{figIntro}). The mean flow $\bar U$  varies \emph{linearly} with~$F$ on both the lower and the upper branch. On the upper branch, form stress $\fs$ is negligible and $U\approx F/\mu$. On the lower branch the forcing~$F$ is weak and the dynamics is linear.  Furthermore, for both small and large $\beta_*$ the transition regime between the upper and lower branches is terminated by a ``\emph{drag crisis}'' at which the form stress abruptly vanishes and the system jumps discontinuously to the upper branch. The lower and upper branches, and the drag crisis, are largely  anticipated by results from low-order truncated models.

A main novelty here, associated with geostrophic turbulence, is the phenomenology of the transition regime:  the lower branch flow becomes unstable at a critical value of~$F$.  Further increase of~$F$ above the critical value results in transient eddies and active geostrophic turbulence. The turbulent transition regime is qualitatively different for the  two values of $\bs$ in  figure~\ref{figIntro}.  For open geostrophic contours (large $\bs$) the flow is  homogeneously distributed over the domain and $\bar U$ is almost constant as the forcing $F$ increases. For closed geostrophic contours (small $\bs$) the flow is spatially inhomogeneous and  is channeled into a  narrow boundary layers separating almost stagnant ``dead zones''; in this case $\bar U$ continues to vary roughly linearly with $F$. The representative transition-regime solutions indicated in figure~\ref{figIntro} are discussed further in sections~\ref{sec:example_closed} and~\ref{sec:example_open}.

The  insensitivity of time-mean large-scale flow $\bar U$  to the strength of the wind stress  $F$ for the large-$\beta$ case in figure~\ref{figIntro} is reminiscent of the ``\textit{eddy saturation}'' phenomenon   identified in eddy-resolving models of the Southern Ocean \citep{Hallberg-Gnanadesikan-2001,Tansley-Marshall-2001,Hallberg-Gnanadesikan-2006,Hogg-etal-2008,Nadeau-Straub-2009,Farneti-etal-2010,Nadeau-Straub-2012,Meredith-etal-2012,Morisson-Hogg-2013,Munday-etal-2013,Farneti-coauthors-2015,Nadeau-Ferrari-2015}. Indications of eddy saturation appear also in observations of the Southern Ocean \citep{Boning-etal-2008,Firing-etal-2011,Hogg-etal-2015}. Eddy saturaton has been of great interest because there is an observed  trend in increasing strength of the westerly winds over the Southern Ocean \citep{Thompson-Solomon-2002,Marshall-2003,Swart-Fyfe-2012}, raising the question of how the  transport of the Antarctic Circumpolar Current will change.  \cite{Straub-1993} first predicted that  transport should become insensitive to the wind stress forcing at sufficiently high wind stress. However, Straub's argument invoked baroclinicity and channel walls as crucial ingredients for eddy saturation. Following Straub, most previous explanations of eddy saturation argue that transport is linearly proportional to isopycnal slopes, and those slopes have a hard maximum set by the marginal condition for baroclinic instability.  Thus we are surprised here to discover that a single-layer fluid in a doubly-periodic geometry exhibits impressive eddy saturation: in figure~\ref{figIntro} the time mean large-scale flow $\bar U$ only doubles as $\Fs$ varies from $0.2$ to~30. We discuss this ``barotropic eddy saturation'' further in section~\ref{sec:saturation} and we speculate on its relation to the baroclinic eddy saturation  exhibited by Southern  Ocean  models in the conclusion  section \ref{conclusion}.

\section{Formulation \label{sec:form}}

We consider barotropic flow in a beta-plane fluid layer with depth $H-h(x,y)$, where $h(x,y)/H$ is order Rossby number. The fluid velocity consists of a large-scale zonal flow, $U(t)$, along the $x$-axis plus smaller scale eddies with velocity $(u,v)$; thus the total flow is
\beq
\bU\defn\(\, U(t)+u(x,y,t)\,,\,v(x,y,t)\,\)\per
\label{eq:deftotU}
\eeq
The eddying component of the flow is derived from an eddy streamfunction $\psi(x,y,t)$ via $(u,v) = (-\psi_y, \psi_x)$; the total streamfunction is $ -U(t) y + \psi(x,y,t)$ with the large-scale flow  $U(t)$ evolving as in \eqref{eq:U_t}. The relative vorticity is $\zeta =   \psi_{xx}+ \psi_{yy}$, and the QGPV is
\beq
f_0+ \beta y + \underbrace{\zeta +\h}_{\defn q}\per \label{beq11}
\eeq
In~\eqref{beq11}, $f_0$ is the Coriolis parameter in the center of the domain, $\beta$ is the meridional planetary vorticity gradient and $\h(x,y)= f_0 h(x,y)/H$ is the topographic contribution to potential vorticity  or simply the \emph{topographic PV}.
The QGPV equation is:
\beq
q_t + \J(\psi - U y,q+\beta y)+ \D \zeta = 0\com
\label{eq:z_NL}
\eeq
where $\J$ is the Jacobian, $\J(a,b)\defn a_x b_y -a_y b_x$. With Navier--Stokes viscosity $\nu$ and linear Ekman drag $\mu$ the ``dissipation operator" $\D$ in~\eqref{eq:z_NL} is
\beq
\D\defn\mu - \nu\lap\per
\label{Ddef}
\eeq
The domain is periodic in both the zonal and meridional direction, with size $2 \upi L \times 2 \upi L$. In numerical solutions, instead of Navier--Stokes viscosity $\nu \lap$ in~\eqref{Ddef}, we  use either hyperviscosity $\nu_4 \nabla^8$, or a high-wavenumber filter. Thus we achieve a regime in which  the role of lateral dissipation is limited to removal small-scale vorticity: the lateral dissipation has a very small effect on larger scales and energy dissipation is mainly due to Ekman drag~$\mu$. Therefore we generally neglect $\nu$ except when discussing the enstrophy balance, in which $\nu$ is an important sink.


The energy and enstrophy of the fluid are defined through:
\begin{align}
E \defn \underbrace{\half U^2}_{\defn E_U} + \underbrace{\half \la|\grad\psi|^2\ra}_{\defn E_{\psi}}\andd Q \defn \underbrace{\beta U}_{\defn Q_U} + \underbrace{\half \la q^2\ra}_{\defn Q_{\psi}}\per\label{eq:E_Q_defs}
\end{align}
Appendix~\ref{app:balances} summarizes the energy and enstrophy balances among the various flow components.

The model formulated in~\eqref{eq:U_t} and~\eqref{eq:z_NL} is the simplest process model which can used to investigate homogeneous beta-plane turbulence driven by a large-scale  wind stress applied at the surface of the fluid.

Although the forcing $F$ in~\eqref{eq:U_t} is steady, the solution often is not: with strong forcing  the flow  spontaneously develops  time-varying eddies. In those cases it is useful to decompose  the eddy streamfunction $\psi$  into time-mean ``standing eddies'', with streamfunction $\bar \psi$,  and  residual ``transient eddies'' $\psi'$:
\beq
\psi(x,y,t) = \bar \psi(x,y) + \psi'(x,y,t)\per \label{tStand}
\eeq
All  fields can then  be decomposed  into time-mean and transient components e.g., $U(t) = \bar U + U'(t)$. A main question is how $\bar U$ depends on $F$, $\mu$, $\beta$, as well as the statistical and geometrical properties of the topographic~PV~$\h$.

The form stress $\fs$ in~\eqref{eq:U_t}  necessarily acts as increased frictional drag on the large-scale mean flow $U$. This becomes apparent from the energy balance of the eddy field, which is obtained through $\overline{\la-\psi\times\textrm{\eqref{eq:z_NL}}\ra}$:
\beq
\overline{U\la \psi \h_x\ra } =   \overline { \laa \mu |\grad\psi|^2+ \nu \zeta^2\raa}\per\label{eq:U_fs}
\eeq
The right hand side of~\eqref{eq:U_fs} is positive definite and thus $U(t)$ is positively correlated with the form stress $\la \psi \h_x\ra$ i.e., on average the topographic form stress acts as an increased drag on the large-scale flow $U$.

\section{ Topography, parameter values and illustrative  solutions \label{examples}}

Although the barotropic quasi-geostrophic model summarized in section~\ref{sec:form} is idealized, it is instructive to estimate $U$ using numbers loosely inspired by the dynamics of the Southern Ocean: see table~\ref{tab:SOvalues}. Without form stress, the equilibrated large-scale velocity obtained from the large-scale momentum equation~\eqref{eq:U_t} using $F$ from table~\ref{tab:SOvalues} is
 \beq
 F\big/\mu= 0.77\,\mathrm{m}\,\mathrm{s}^{-1}\per\label{Fovermu}
 \eeq
The main point of \cite{Munk-Palmen-1951} is that this drag-balanced large-scale velocity is far too large. For example, the implied transport through a meridional section $1000\,\textrm{km}$ long is over $3\times 10^9\,\textrm{m}^3\,\textrm{s}^{-1}$; this is larger by a factor of about twenty than the observed transport of the  Antarctic Circumpolar Current \citep{Koenig-etal-2016,Donohue-etal-2016}.


\renewcommand{\arraystretch}{1.35}
\begin{table}
	\vspace{-1em}
\caption{Numerical values characteristic of the Southern Ocean; $f_0=-1.26\times10^{-4}\,\textrm{s}^{-1}$ and $\beta=1.14\times10^{-11}\,\textrm{m}^{-1}\textrm{s}^{-1}$. The drag coefficient $\mu$ is taken from \citet{Arbic-Flierl-2004}.}
\vspace{-1em}
\begin{center}
\begin{tabular}{l c l} \\
domain size, $2\upi L\times 2 \upi L$ & $L$ & $800\,\textrm{km}$ \\
mean depth & $H$ & $4\,\textrm{km}$ \\
density of seawater & $\rho_0$ & $1035\,\textrm{kg}\,\textrm{m}^{-3}$ \\
r.m.s.~topographic height & $h_{\mathrm{rms}}$ & $200\,\textrm{m}$  \\
r.m.s.~topographic PV & $\etarms= f_0 h_{\mathrm{rms}}/H$ \quad & $6.30\times 10^{-6}\,\textrm{s}^{-1}$ \\
Ekman drag coefficient & $\mu$ & $6.30\times 10^{-8}\,\textrm{s}^{-1}$ \\
wind stress & $\tau$ & $0.20\,\textrm{N}\,\textrm{m}^{-2}$ \\
forcing on the right of~\eqref{eq:U_t}& $\wind = \tau/(\rho_0 H)$ \quad & $4.83\times 10^{-8}\,\textrm{m}\,\textrm{s}^{-2}$ \\
topographic length scale & $\leta = 0.0690 L$ & $55.20\,\textrm{km}$ \\
r.m.s.~topographic slope & $h_{\mathrm{rms}}/\leta$ &$3.62 \times 10^{-3}$ \\
a  velocity scale & $\beta \leta^2$ &$3.47\times 10^{-2}\,\textrm{m}\,\textrm{s}^{-1}$ \\
non-dimensional $\beta$ & $\bs=\beta\leta/\etarms$ &$ 1.00\times10^{-1}$  \\
non-dimensional drag & $\mus=\mu/\etarms$   \quad & $ 1.00\times 10^{-2}$\\
non-dimensional forcing & $\Fs=F/(\mu\etarms\leta)$ \quad & $2.20$
\end{tabular}
\end{center}
\label{tab:SOvalues}
\end{table}

\subsection{The topography \label{topoSec}}

If the topographic height has a root mean square  value of order $200\,\textrm{m}$, typical of abyssal hills \citep{Goff-2010}, then $\etarms^{-1}$ is less than 2 days. Thus even rather small topographic features produce a topographic PV with a time scale that is much less than that of the typical drag coefficient in table~\ref{tab:SOvalues}. This order-of-magnitude estimate indicates that the form stress is likely to be large. To say more about form stress we must introduce the model  topography with more detail.

\begin{figure}
\centering
\vspace{-1em}
\includegraphics[width = 0.75\textwidth,trim=0 0 0 8mm,clip]{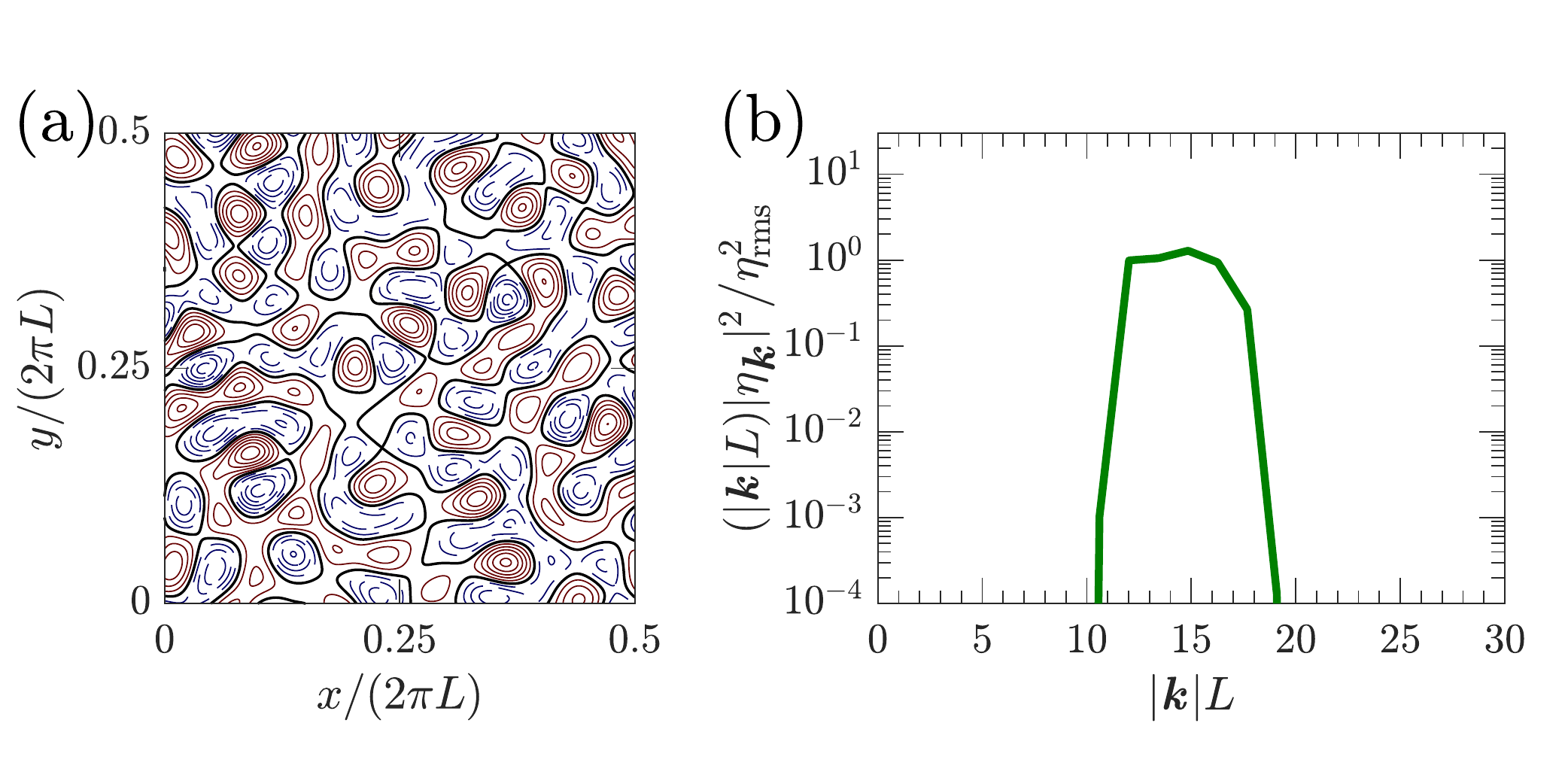}
\vspace{-2em}\caption{The structure and spectrum of the topography used in this study.  Panel (a) shows the structure of the topography for a quarter of the full domain. Solid curves are positive contours, dashed curves negative contours and the thick curves marks the zero contour. Panel (b) shows the 1D power spectrum. The topography has power only within the annulus $12\le |\bk|L \le 18$.}\label{fig:topo}
\end{figure}

The topography is synthesized as $\h(x,y) = \sum_{\bk} \ee^{\ii \bk \bcdot \bx} \h_{\bk}$, with random phases for $\eta_{\bk}$. We consider a homogeneous and isotropic topographic model  illustrated in figure~\ref{fig:topo}. The topography is constructed by confining the wavenumbers $\h_{\bk}$  to a relatively narrow annulus with $12\leq |\bk|L \leq 18$. The spectral cut-off is tapered smoothly to zero at the edge of the annulus. In addition to being homogeneous and isotropic, the topographic model in figure~\ref{fig:topo} is approximately \textit{monoscale} i.e., the topography is characterized by a single length scale determined, for instance, by the central wavenumber $|\bk| \approx 15/L$ in figure~\ref{fig:topo}(b). To assess the validity of the monoscale approximation we characterize the topography using the length scales
\beq
\leta \defn  \sqrt{ \laa\eta^2 \raa \big\slash \laa |\grad \eta|^2 \raa} \andd \Leta \defn  \sqrt{ \laa \big|\grad \nabla^{-2} \eta \big|^2\raa  \big \slash   \laa \eta^2 \raa } \per
\eeq
For the model in figure~\ref{fig:topo}:
\beq
\leta  = 0.0690\,L \andd \Leta = 0.0707\,  L\per
\eeq
(Recall the domain is $2 \upi L \times 2 \upi L$.) Because $\leta\approx\Leta$ we conclude that the topography in figure~\ref{fig:topo} is monoscale to a good approximation and we  use the slope-based length $\leta$ as the  typical length scale of the  topography.

\begin{figure}
\centering
\includegraphics[width = .99\textwidth,trim={0cm 0cm 0cm 13mm},clip]{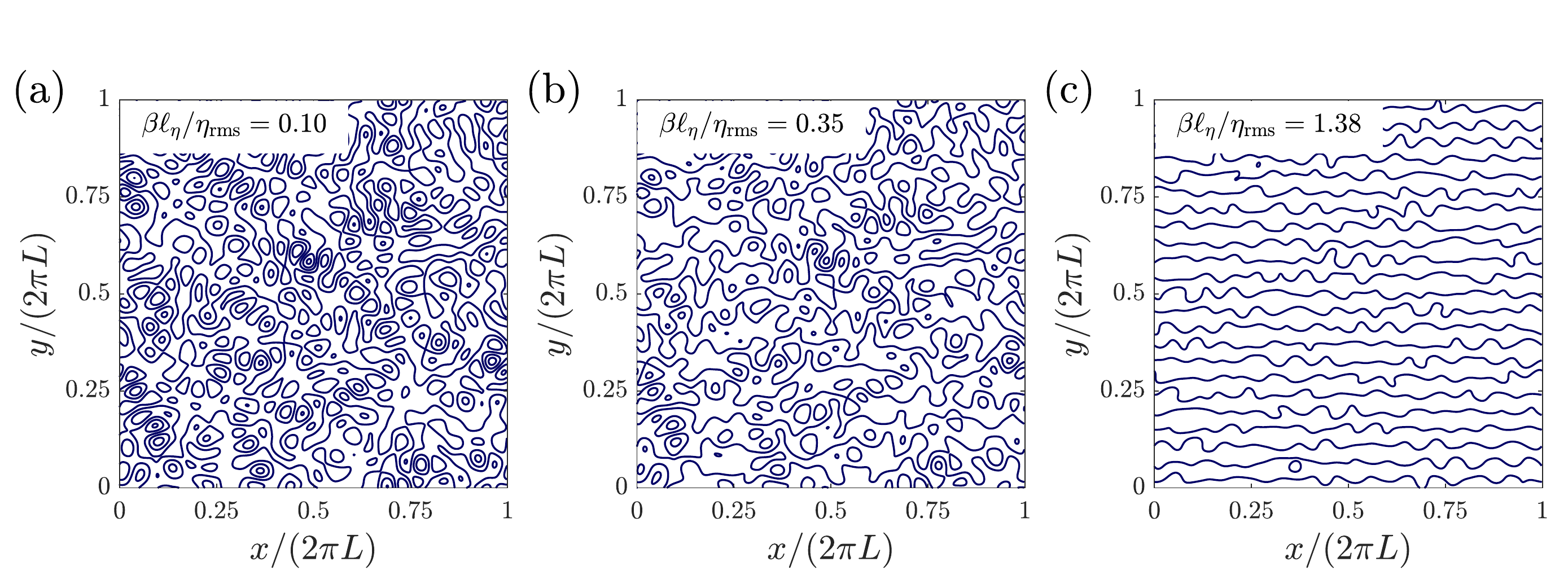}
\vspace{-1em}
\caption{The structure of the geostrophic contours, $\beta y + \eta$, for the monoscale topography of figure~\ref{fig:topo} and for various values of (a)~$\bs=0.10$, (b)~$\bs=0.35$ and (c)~$\bs=1.38$. It is difficult to visually distinguish the geostrophic contours with $\bs=0$ from those with $\bs=0.1$ in panel~(a).}
\label{fig:envPV}
\end{figure}

The isotropic homogeneous monoscale model adopted here has no claims to realism. However, the monoscale assumption greatly simplifies many aspects of the problem because all relevant second-order statistical characteristics of the model topography can be expressed in terms of the two dimensional quantities $\etarms$ and $\leta$ e.g., $\la (\nabla^{-2} \eta_x)^2 \ra = \half\leta^2 \etarms^2$. The main advantage of monoscale topography is that despite the simplicity of its spectral characterization it exhibits the crucial distinction between open and closed \textit{geostrophic contours}: see figure~\ref{fig:envPV}.

\subsection{Non-dimensionalization}\label{sec:nondim}

There are four time scales in the problem: the topographic PV $\etarms^{-1}$, the dissipation $\mu^{-1}$, the period  of topographically excited Rossby waves $(\beta\leta)^{-1}$, and the advective time-scale associated with the forcing  $\leta\mu/F$. From these four time scales we construct the three main non-dimensional  control parameters:
\beq
\mus\defn \mu \big/ \etarms \com \quad  \bs\defn \beta \leta\big/ \etarms \quad\text{and}\quad \Fs \defn  \wind \big/ (\mu\etarms\leta)\per
\label{eq:nondimparams}
\eeq
The parameter $\bs$ is the ratio of the planetary vorticity gradient over the r.m.s.~ topographic PV gradient. There is a fourth parameter $L/\leta$ that measures the scale separation between the domain and the topography. We assume that as $L/\leta \to \infty$ there is a regime of statistically homogeneous two-dimensional turbulence. In other words, as $L/\leta \to \infty $, the flow becomes asymptotically independent of $L/\leta$ so that the large-scale flow $\bar U$ and other statistics, such as $E_{\bar\psi}$, are independent of the domain size $L$.

Besides the  control parameters above, additional parameters are required to characterize the topography. For example, in the case of a multi-scale topography the ratio $L_{\eta}/\leta$ characterizes the spectral width of the power-law range. A main simplification of the monoscale case used throughout this paper is that  we do not have to contend with these additional topographic parameters.

\subsection{Geostrophic contours}

We refer to contours of constant $\beta y + \eta$ as the \emph{geostrophic contours}.  Closed geostrophic contours enclose  isolated pools within the domain --- see figure~\ref{fig:envPV}(a) ---  while open contours thread through the domain in the zonal direction, connecting one side to the other --- see figure~\ref{fig:envPV}(c). The transition between the two limiting cases is controlled by $\bs$. Figure~\ref{fig:envPV}(b) shows an intermediate case with a mixture of closed and open geostrophic contours.

%

It is instructive to consider  the extreme case $\beta=0$. Then  only the geostrophic contour $\eta=0$ is open and all other geostrophic contours are closed. This intuitive conclusion relies on a special property of the random  topography  in figure~\ref{fig:topo}: the topography $-\h$ is statistically equivalent $+\h$. In other words, if $\h(x,y)$ is in the ensemble then so is $-\h(x,y)$.  For further discussion of this conclusion see the discussion of continuum percolation by  \citet{Isichenko-1992}.

If $\beta$ is non-zero but small, in the sense that $\bs\ll1$, then most of the domain is within closed contours: see figure~\ref{fig:envPV}(a). The planetary PV gradient $\beta$ is too small relative to $\grad \eta$ to destroy local pools of closed geostrophic contours.  But $\beta$ dominates the long-range structure of the geostrophic contours and opens up narrow channels of open geostrophic contours.

The other extreme is $\bs \gg 1$;  in this case,  illustrated in  figure~\ref{fig:envPV}(c), all geostrophic contours are open.  Because of its geometric simplicity  the situation with  $\bs \gg 1$  is the easiest to analyze and understand. Unfortunately, the difficult case in figure~\ref{fig:envPV}(a) is most relevant to ocean conditions. In sections~\ref{sec:example_closed} and~\ref{sec:example_open} we illustrate  the two  cases using numerical solutions of~\eqref{eq:U_t} and~\eqref{eq:z_NL}.

\subsection{An example with mostly closed geostrophic contours: $\bs =0.1$ \label{sec:example_closed}}

\begin{figure}
\centering
\includegraphics[width = .8\textwidth]{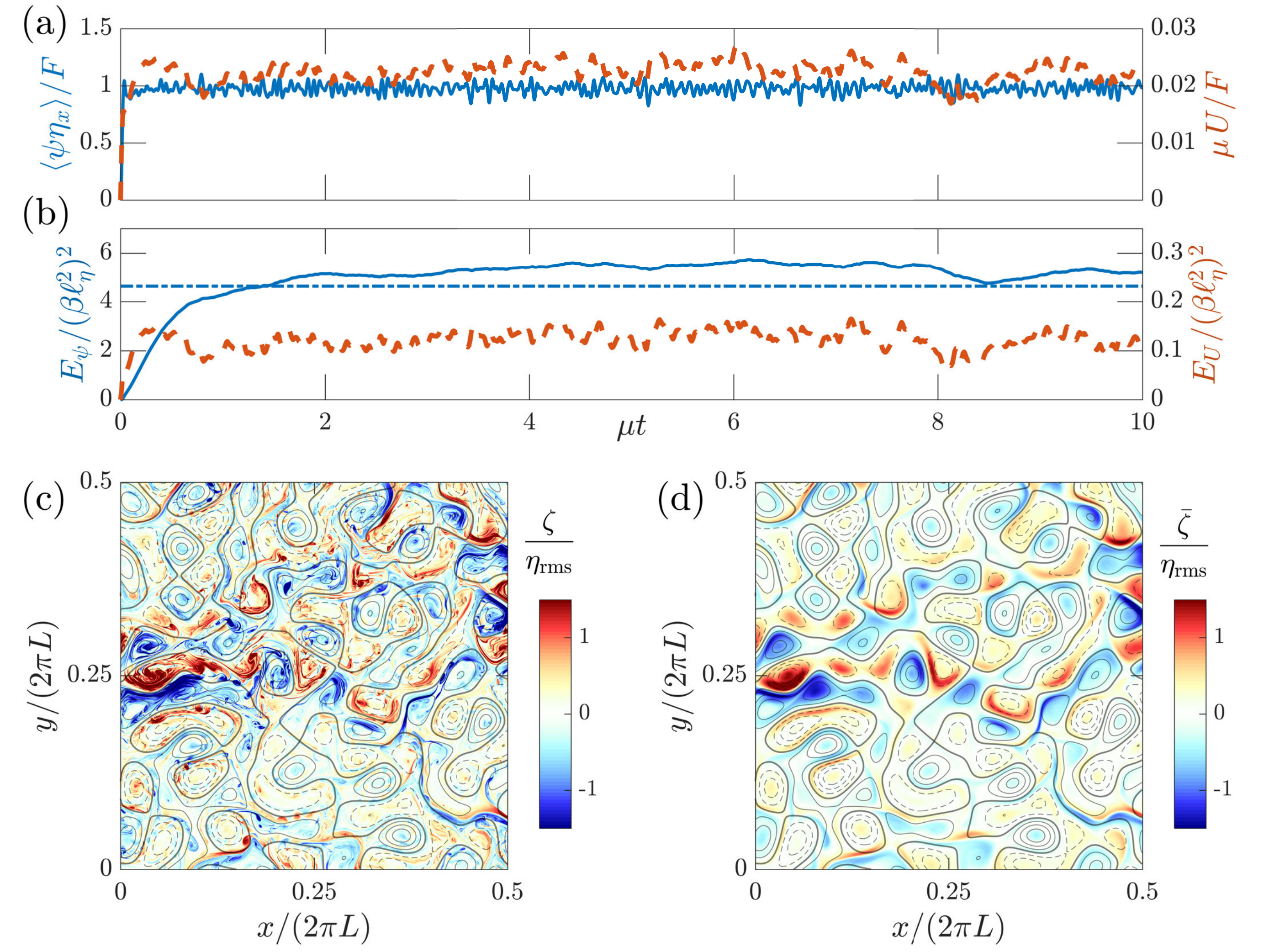}
\vspace{-1em}\caption{A solution with closed geostrophic contours: $\bs=0.1$, $\Fs=2.20$ and $\mus=10^{-2}$. Panel~(a) shows the evolution of the large-scale zonal  flow $U(t)$ (dashed) and the form stress $\la \psi \eta_x \ra$ (solid). Panel (b) shows the evolution of $E_\psi$ (solid) and $E_U$ (dashed). The dash-dot line in panel~(b) is the energy level of the standing eddies, $E_{\bar\psi}=\half\la |\grad \bar \psi|^2\ra$. Panel~(c) shows a snapshot of the relative vorticity, $\zeta$, (shaded) at $\mu t=10$ in one-quarter of the domain overlying the topographic PV (solid contours are positive $\eta$ and dashed contours are negative). Panel~(d) shows the time-mean $\bar{\zeta}$. A movie showing the evolution of $q=\z+\h$ and $\psi-Uy$ from rest is found in \textbf{Supplementary Materials}.\label{fig:eg_closed_A}}

\centering
\includegraphics[width = \textwidth]{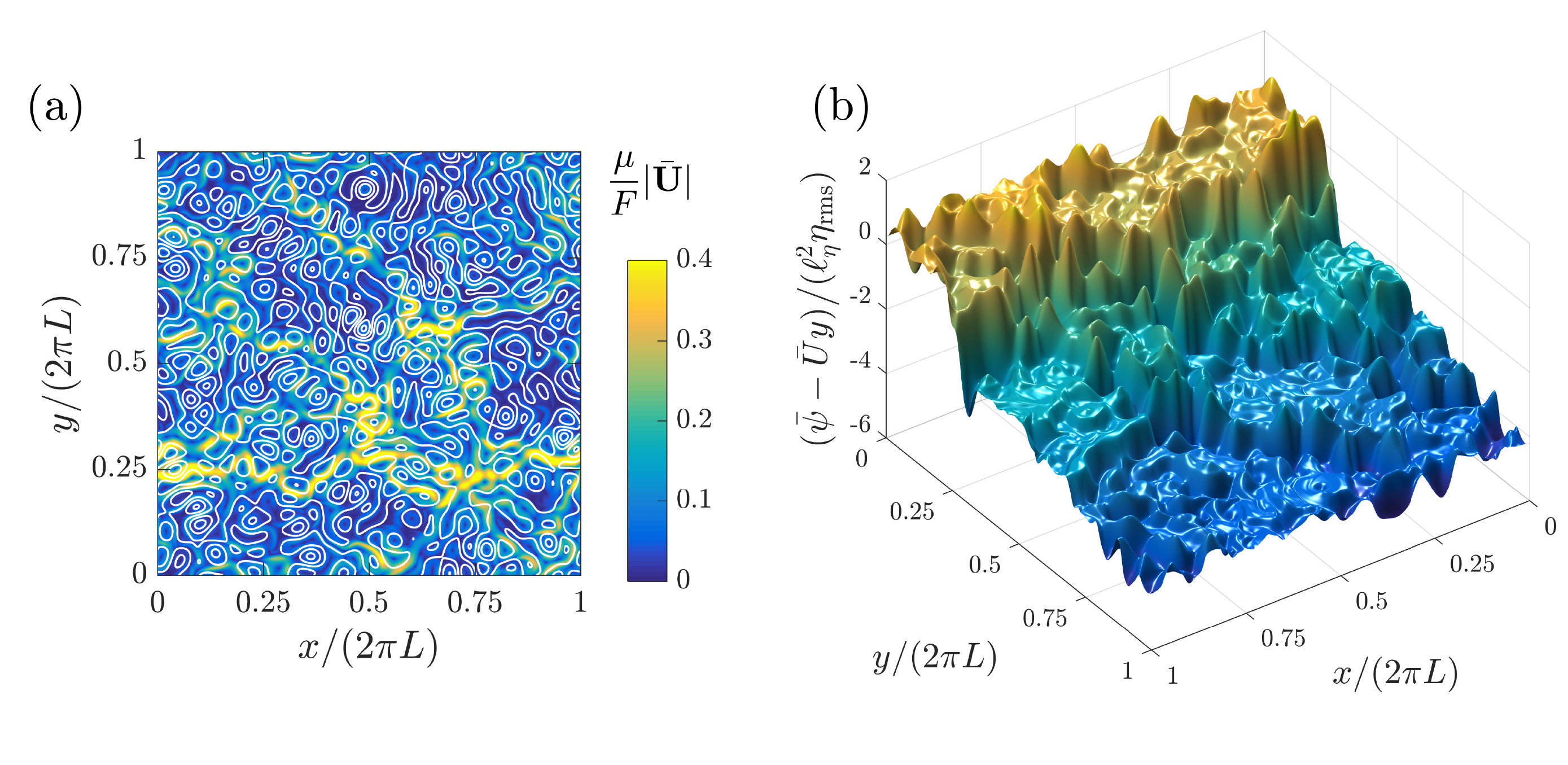}
\vspace{-3em}\caption{A solution with $\bs=0.1$, $\Fs=2.20$ and $\mus=10^{-2}$. (a)~The speed of the  time-mean flow, $|\bar{\bU}|$ is indicated by colors;  the geostrophic contours $\beta y+\h$  are shown as white curves. (b)~Surface plot of the total time-mean streamfunction, $\bar{\psi}-\bar{U}y$. }\label{fig:eg_closed_B}
\end{figure}

Figures~\ref{fig:eg_closed_A} and~\ref{fig:eg_closed_B} show a numerical solution for a case with mostly closed geostrophic contours; this is  the $\bs=0.1$ ``boxed'' point indicated in figure \ref{figIntro}. In this illustration we use the Southern Ocean parameter values in table~\ref{tab:SOvalues}. The solution employs $1024 \times 1024$ grid points with a high-wavenumber filter that removes vorticity at small scales. The system is evolved using the ETDRK4 time-stepping scheme of \citet{Cox-Matthews-2002} with the refinement of \citet{Kassam-Trefethen-2005}.

Figure~\ref{fig:eg_closed_A}(a) shows the evolution of the large-scale flow $U(t)$ and the form stress $\la \psi \eta_x\ra$. After a spin-up of duration $\sim\mu^{-1}$ the flow achieves a statistically steady state in which  $U(t)$ fluctuates around the time mean $\bar U$. In figure~\ref{fig:eg_closed_A}(a) the form stress $\fsb$ balances almost 98\% of $F$,  so that $U(t) $ is very much smaller than $F/\mu$ in~\eqref{Fovermu}. The time-mean of the large-scale flow is
$
\bar U =1.70\, \textrm{cm}\,\textrm{s}^{-1}\com
$
which is 2.2\% of the velocity $F/\mu$ in~\eqref{Fovermu}. Figure~\ref{fig:eg_closed_A}(b) shows the evolution of the energy: the eddy energy $E_{\psi}$ is about 50 times greater than the large-scale energy $E_U$. With the decomposition of $\psi$ in~\eqref{tStand} the time-mean eddy energy $\overline{E_{\psi}}$ is decomposed into $E_{\bar \psi}+\overline{E_{\psi'}}$; the dash-dot line in figure~\ref{fig:eg_closed_A}(b) is the energy of the standing component $E_{\bar \psi}$: the transient eddies are less energetic than the standing eddies. This is also evident by comparing the snapshot of the relative vorticity $\z$ in figure~\ref{fig:eg_closed_B}(c) with the time mean $\bar \zeta$ in figure~\ref{fig:eg_closed_B}(d): many features in the snapshot are also seen in the time mean.

Figures~\ref{fig:eg_closed_A}(c) and~(d) show that there is anti-correlation between the time-mean relative vorticity and the topographic~PV: $\corr(\bar\zeta,\eta) = - 0.53$, where the correlation between two fields $a$ and $b$  is
\beq
	\corr(a,b) \defn \la a  b \ra \Big/ \sqrt{\la a^2\ra \la b^2 \ra} \per
\eeq
Another statistical characterization of the solution is that $\corr(\bar\psi,\eta_x)=0.06$, showing  that  a  weak correlation between the standing  streamfunction $\bar \psi$ and the topographic~PV gradient $\eta_x$ is sufficient to produce a form stress balancing about $98$\% the applied wind stress.

The most striking characterization of the time-mean flow is that it is very weak in most of the domain: figure~\ref{fig:eg_closed_B}(a) shows that most of the flow through the domain is channeled into a relatively narrow band centered very roughly on $y/(2\upi L) = 0.25$: this ``main channel" coincides with the extreme values of $\zeta$ and $\bar \zeta$ evident in figures~\ref{fig:eg_closed_A}(c) and~(d) (notice that figures~\ref{fig:eg_closed_A}(c) and~(d) show only a quarter of the flow domain). Outside of the main channel the time-mean flow is weak. We emphasize that $\bar{U}=1.70\, \textrm{cm}\,\textrm{s}^{-1}$ is an unoccupied mean that is not representative of the larger velocities in the main channel: the channel velocities are 40 to 50 times larger than $\bar U$.

Figure~\ref{fig:eg_closed_B}(b) shows the streamfunction $\bar \psi(x,y)- \bar U y$ as a surface above the $(x,y)$-plane. The mean streamfunction surface appears as a terraced hillside: the mean slope of the hillside is $-\bar U$ and  stagnant pools, with constant $\bar{\psi}-\bar{U} y$, are the flat terraces carved into the hillside. The existence of these stagnant dead zones is explained by the closed-streamline theorems of \citet{Batchelor-1956} and  \citet{Ingersoll-1969}. The dead zones are separated by boundary layers and the strongest of these boundary layers is the main channel which appears as the large cliff located roughly at $y/(2\upi L)=0.25$ in figure~\ref{fig:eg_closed_B}(b). The main channel is determined by a narrow band of geostrophic contours that are opened by the small $\beta$-effect: this provides an open path for flow through the  disordered topography.

 \subsection{An example with open geostrophic contours: $\bs=1.38$ \label{sec:example_open}}

 \begin{figure}
 \centering
 \includegraphics[width = .8\textwidth]{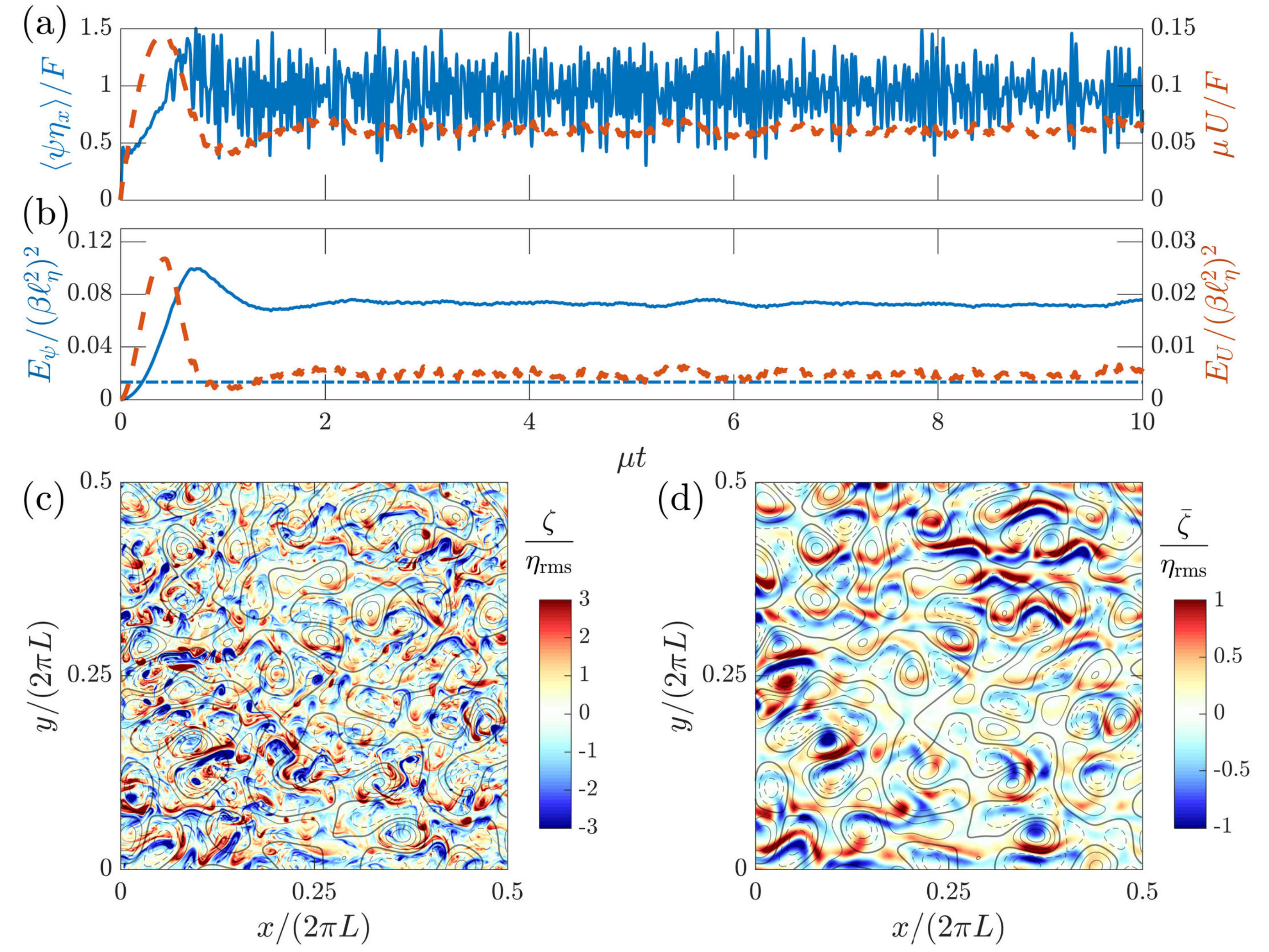}
 \vspace{-1em}\caption{A solution with $\bs=1.38$  and  open geostrophic contours. All other parameters as in figure~\ref{fig:eg_closed_A}. Panels also as in figure~\ref{fig:eg_closed_A}. Note that the color scale is different between panels~(c) and~(d). A movie showing the evolution of $q=\z+\h$ and $\psi-Uy$ from rest can be found in \textbf{Supplementary Materials}.\label{fig:eg_open_A}}
%
%
%
\vspace{-.3em}
 \centering
 \includegraphics[width = \textwidth,trim={0cm 0cm 0cm 5.5mm},clip]{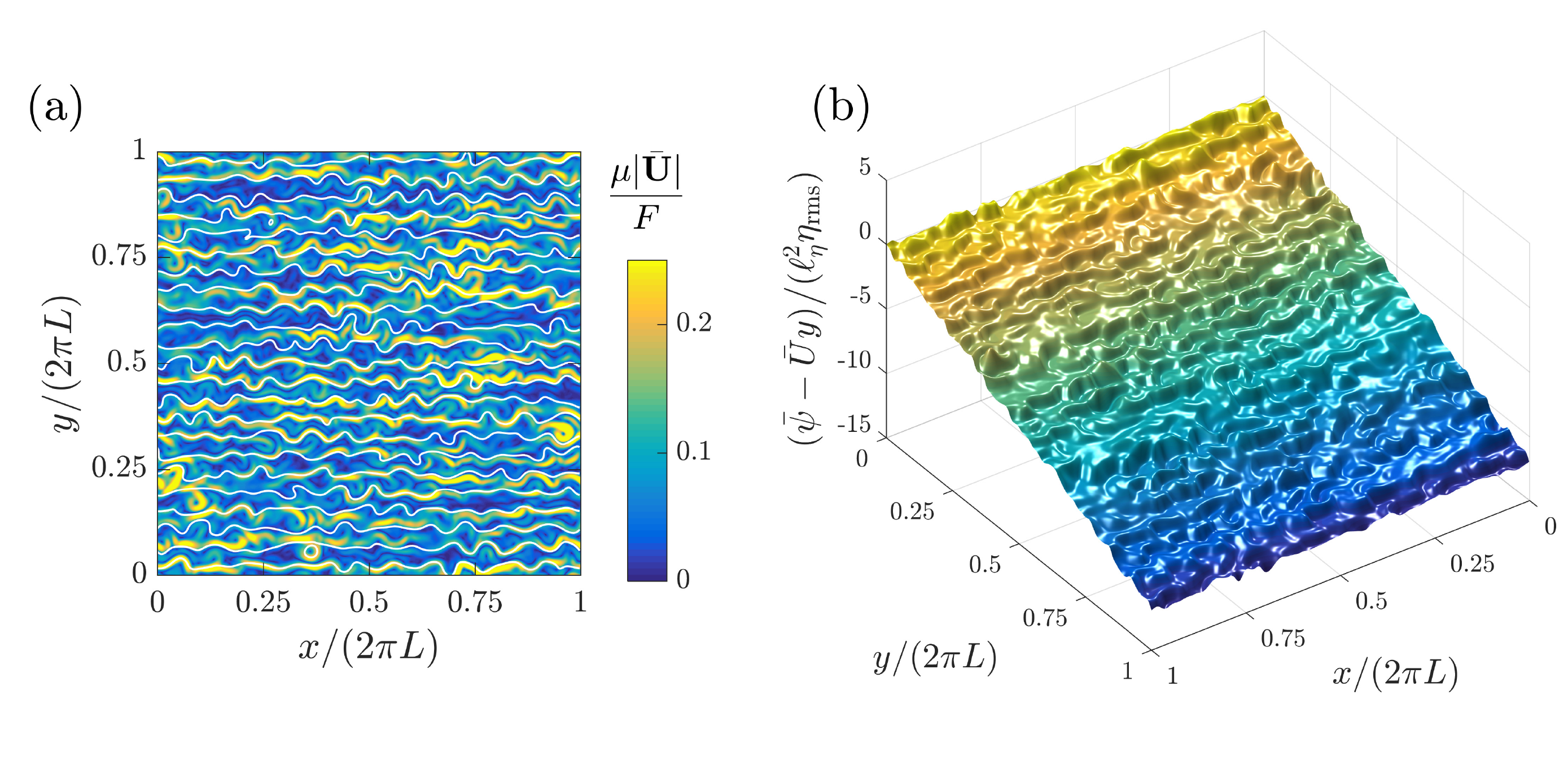}
 \vspace{-3.5em}\caption{A solution with $\bs=1.38$. All other parameters are as in figure~\ref{fig:eg_closed_B}.  (a) The  speed of the  time-mean flow, $|\bar{\bU}|$ (colors);  the geostrophic contours $\beta y+\h$  are shown as white curves.  (b) Surface plot of the total time-mean streamfunction, $\bar{\psi}-\bar{U}y$. \label{fig:eg_open_B}}
 \end{figure}

 Figures~\ref{fig:eg_open_A} and~\ref{fig:eg_open_B} show a solution for the same parameters as in~Figures~\ref{fig:eg_closed_A} and~\ref{fig:eg_closed_B}, except $\bs=1.38$; this is  the $\bs=1.38$ ``boxed'' point indicated in figure \ref{figIntro}. The geostrophic contours are open throughout the domain. The most striking difference when compared to the previous blocked case in section~\ref{sec:example_closed} is that there are no dead zones; the flow is more evenly spread throughout the domain: compare figure~\ref{fig:eg_open_B} with figure~\ref{fig:eg_closed_B}. The time-mean streamfunction in figure~\ref{fig:eg_open_B}(b) is not ``terraced''. Instead $\bar \psi - \bar U y$ in figure~\ref{fig:eg_open_B}(b) is better characterized as  a bumpy slope.

The large-scale flow is $\bar U =4.68\, \textrm{cm}\,\textrm{s}^{-1}$, which is again very much smaller than the flow that would exist in the absence of topography: $\bar U$ is only 6\% of $F/\mu$. The eddy energy $E_{\psi}$ is roughly 15 times larger than the large-scale flow energy $E_U$. Moreover,  the energy of the transient eddies, shown in figure~\ref{fig:eg_open_A}(b), is in this case much larger than that of the standing eddies. This is also apparent by comparing the instantaneous and time-mean relative vorticity fields in figures~\ref{fig:eg_open_A}(c) and~(d). In anticipation of the discussion in section~\ref{sec:saturation} we remark that these strong transient eddies act as PV diffusion on the time-mean QGPV~\citep{Rhines-Young-1982}.

In contrast to the  example of section \ref{sec:example_closed}, the relative vorticity is positively correlated with the topographic PV: $\corr(\bar\zeta,\eta)=0.23$. Because of the strong transient eddies the sign of $\corr(\bar\zeta,\eta)$ is not apparent by visual inspection of figures~\ref{fig:eg_open_A}(c) and~(d). The form-stress correlation  is $\corr(\bar\psi,\eta_x)=0.15$. Again, this weak correlation is sufficient to produce a form stress balancing 94\% of the wind stress.

\section{Flow regimes and a parameter survey \label{survey}}

In this section we present a comprehensive suite of numerical simulations of~\eqref{eq:U_t} and~\eqref{eq:z_NL} using the topography of figure~\ref{fig:topo}(a). A complete survey of the parameter space is complicated by the existence of at least three control parameters~\eqref{eq:nondimparams}. In the following survey we use
\beq
	\mus=10^{-2} \com
\eeq
and vary the strength of the non-dimensional large-scale wind forcing $\Fs$ and the non-dimensional planetary vorticity gradient~$\bs$. Most the solutions presented use $512^2$ grid points; additionally, a few $1024^2$ solutions were obtained to test sensitivity to resolution (we found none). Unless stated otherwise, numerical simulations are initiated from rest and time-averaged quantities are calculated  by averaging the fields over the interval $10\le \mu t\le30$.

\subsection{Flow regimes: the lower branch, the upper branch, eddy~saturation and the drag crisis}

\begin{figure}
\centering
\includegraphics[width = .85\textwidth,trim={0cm 0cm 0cm 0mm},clip]{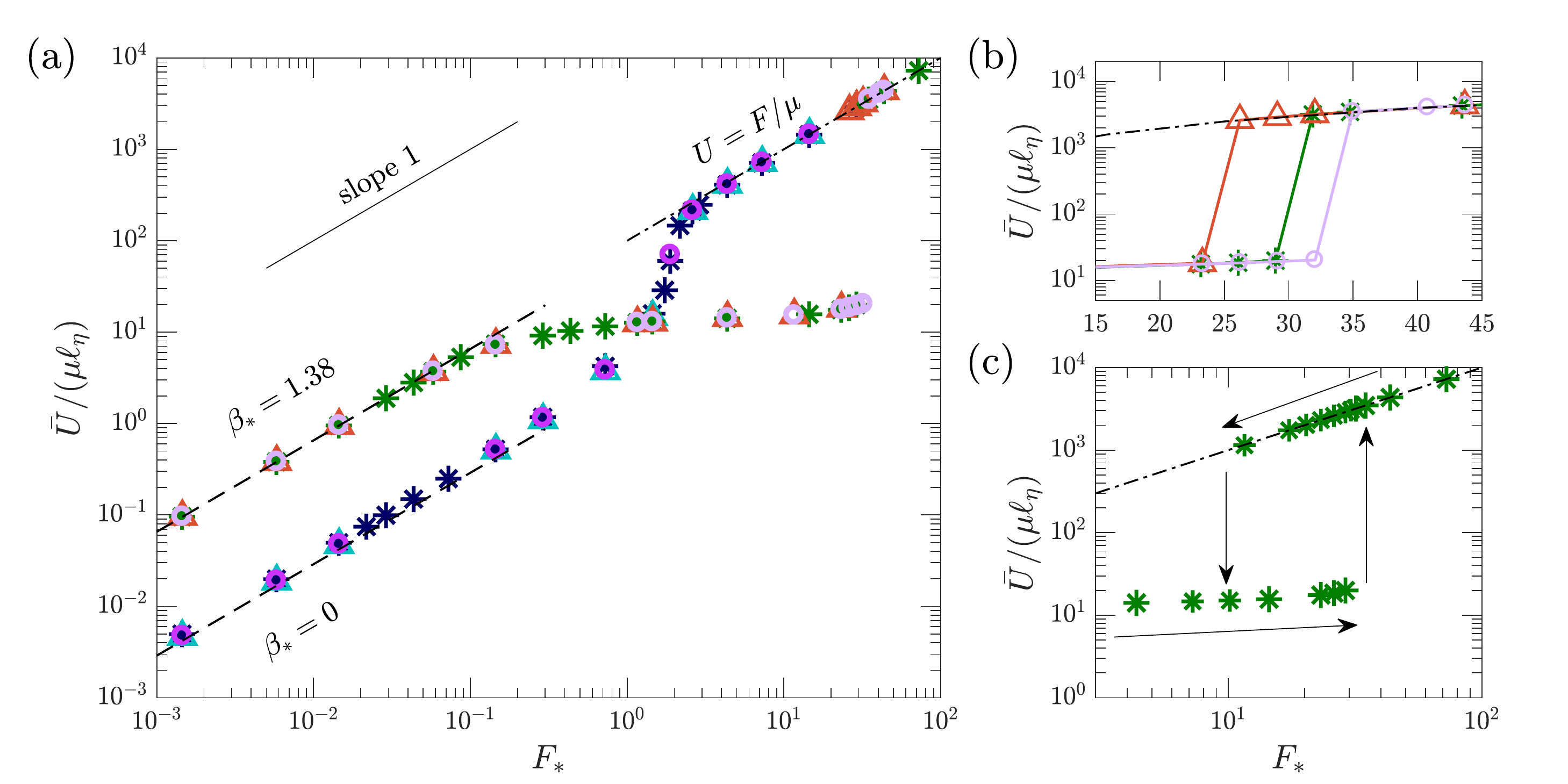}
\vspace{-0.5em}\caption{(a) The equilibrated large-scale mean flow $\bar{U}$ as a function of $\Fs$ for cases with $\bs=0$ and $\bs=1.38$. Shown are results for three different monoscale topography realizations (each denoted with a different marker symbol: \textasteriskcentered, {\small$\triangle$}, $\circ$) all with the spectrum in figure~\ref{fig:topo}(c). Other parameters are in Table~\ref{tab:SOvalues} e.g., $\mus =10^{-2}$. Panel (b) shows a detailed view of the transition from the lower to the upper branch solution for the case with  $\bs=1.38$ and panel (c) shows the hysteretic solutions for one of the topography realizations with $\bs=1.38$. Dashed lines in panel (a) correspond to asymptotic expressions derived in section~\ref{sec:weak} and dash-dotted lines in all panels mark the solution: $U=F/\mu$.}\label{fig:U_F_b0_b20}
\end{figure}

Keeping $\bs$ fixed and increasing the wind forcing $\Fs$ from very small values  we find that the statistically equilibrated solutions show either one of the two characteristic behaviors depicted in figure~\ref{fig:U_F_b0_b20}.

For $\beta=0$, or for values of $\bs$ much less than one, we find that the equilibrated time-mean large-scale flow $\bar{U}$ scales linearly with $\Fs$ when $\Fs$ is very small. On this lower branch the large-scale velocity is
\beq
 \bar U \approx F\big/\mueff\com \qquad \text{with} \qquad \mueff \gg \mu \per
 \label{lowerBranch}
\eeq
In section~\ref{sec:weak} we provide an analytic expression for the effective drag $\mueff$ in~\eqref{lowerBranch}; this analytic expression is shown by the dashed lines in figure~\ref{fig:U_F_b0_b20}. As $\Fs$ increases, $\bar U$  transitions to a different linear relation with
\beq
 \bar{U}\approx F\big/\mu\per \label{upperBranch}
\eeq
On this upper branch the form stress is essentially zero and $\wind$ is balanced by bare drag $\mu$.

For the $\beta=0$ case shown in figure~\ref{fig:U_F_b0_b20} the transition between the lower and upper branch occurs in the range  $0.6< \Fs <3$; the equilibrated $\bar U$ increases by a factor of more than 200 within this interval. On the other hand, for $\bs$ larger than 0.05, we find a quite  different behavior, illustrated in figure~\ref{fig:U_F_b0_b20} by the runs with $\bs=1.38$. On the lower branch $\bar U$ grows linearly with $F$ with a constant $\mueff$ as in~\eqref{lowerBranch}. But the linear increase in $\bar U$ eventually ceases and instead  $\bar U$ then grows at a much more slower rate as $\wind$ increases. For the case $\bs=1.38$ shown in figure~\ref{fig:U_F_b0_b20},  $\bar{U}$ only doubles as $\wind$ is increased over 150-fold from $\Fs=0.2$ to $30$. We identify this regime, in which $\bar U$ is insensitive to changes in $\wind$, with the ``eddy~saturation'' regime of \citet{Straub-1993}. As $\wind$ increases further the flow exits the eddy~saturation regime via a ``drag crisis'' in which the form stress abruptly vanishes and $\bar U$ increases by a factor of over  200 as the solution jumps to the upper branch~\eqref{upperBranch}. In~figure~\ref{fig:U_F_b0_b20} this drag crisis is a discontinuous transition from the eddy~saturated regime to the upper branch. The drag crisis, which requires non-zero $\beta$, is qualitatively different from the continuous transition between the upper and lower branches which is characteristic of flows with  small (or zero) $\bs$.

Figure~\ref{fig:U_F_b0_b20} shows results obtained with three different realizations of monoscale topography viz., the topography illustrated in figure~\ref{fig:topo}(a) and two other realizations with the monoscale  spectrum of figure~\ref{fig:topo}(b). The large-scale flow $\bar U$ is insensitive to these changes in topographic detail; in this sense the large-scale flow is ``self-averaging". However, the  location of the drag crisis depends  on differences between the three realizations: panel~(b) of figure~\ref{fig:U_F_b0_b20} shows that the location of the  jump from lower to upper branch is realization-dependent: the three realizations  jump to the upper branch at  different values of $\Fs$.

The case with $\bs=0.1$, which corresponds a value close to realistic (cf.~Table~\ref{tab:SOvalues}), does show a drag crisis, i.e., a discontinuous  jump from the lower to the upper branch at $\Fs\approx3.9$; see figure~\ref{figIntro}. However, the eddy~saturation regime, i.e., the regime in which $\bar{U}$ grows with wind stress forcing are at rate less than linear, is not nearly as pronounced as in the case with $\bs=1.38$ shown in figure~\ref{fig:U_F_b0_b20}(a).

\subsection{Hysteresis and multiple flow patterns}

Starting  with a severely truncated spectral model of the atmosphere introduced by \citet{Charney-DeVore-79}, there has been considerable interest in the possibility that topographic form stress  might result in multiple stable large-scale flow patterns which might explain blocked and unblocked  states of atmospheric circulation. Focussing on atmospheric conditions, \citet{Tung-Rosenthal-1985} concluded that the results of low-order truncated models are not a reliable guide to the full nonlinear problem: although multiple stable states still exist in the full problem, these occur only in a restricted parameter range that is not characteristic of Earth's atmosphere.

With this meteorological  background in mind, it is interesting that in the oceanographic parameter regime emphasized here, we easily found multiple equilibrium solutions on either side of the drag crisis. After increasing $\wind$ beyond the crisis point, and jumping to the upper branch, we performed additional numerical simulations by decreasing $\wind$ and using initial conditions obtained from the upper-branch solutions at larger values of $F$. Thus we moved down the upper branch, past the crisis, and determined a range of wind stress forcing values with multiple  flow patterns. Panel (c) of figure~\ref{fig:U_F_b0_b20}, with $\bs=1.38$, shows that multiple states co-exist in the range $11\le \Fs\le29$. Note that for quasi-realistic case with $\bs=0.1$ multiple solutions exist only in the limited parameter range $2.9\le \Fs\le 3.9$ . These co-existing flows  differ qualitatively: the lower-branch flows, being near the drag crisis, have an important transient eddy component and almost all of $\wind$ is balanced by form stress: the example discussed in connection with  figures~\ref{fig:eg_closed_A} and~\ref{fig:eg_closed_B} is typical. On the other hand, the co-existing  upper-branch solutions are steady (that is $\psi'=U'=0$) and nearly all of the wind stress is balanced by bottom drag so that $\mu U/F\approx 1$.


\subsection{A survey\label{sec:survey}}

\begin{figure}
\centering
\includegraphics[width = .6\textwidth]{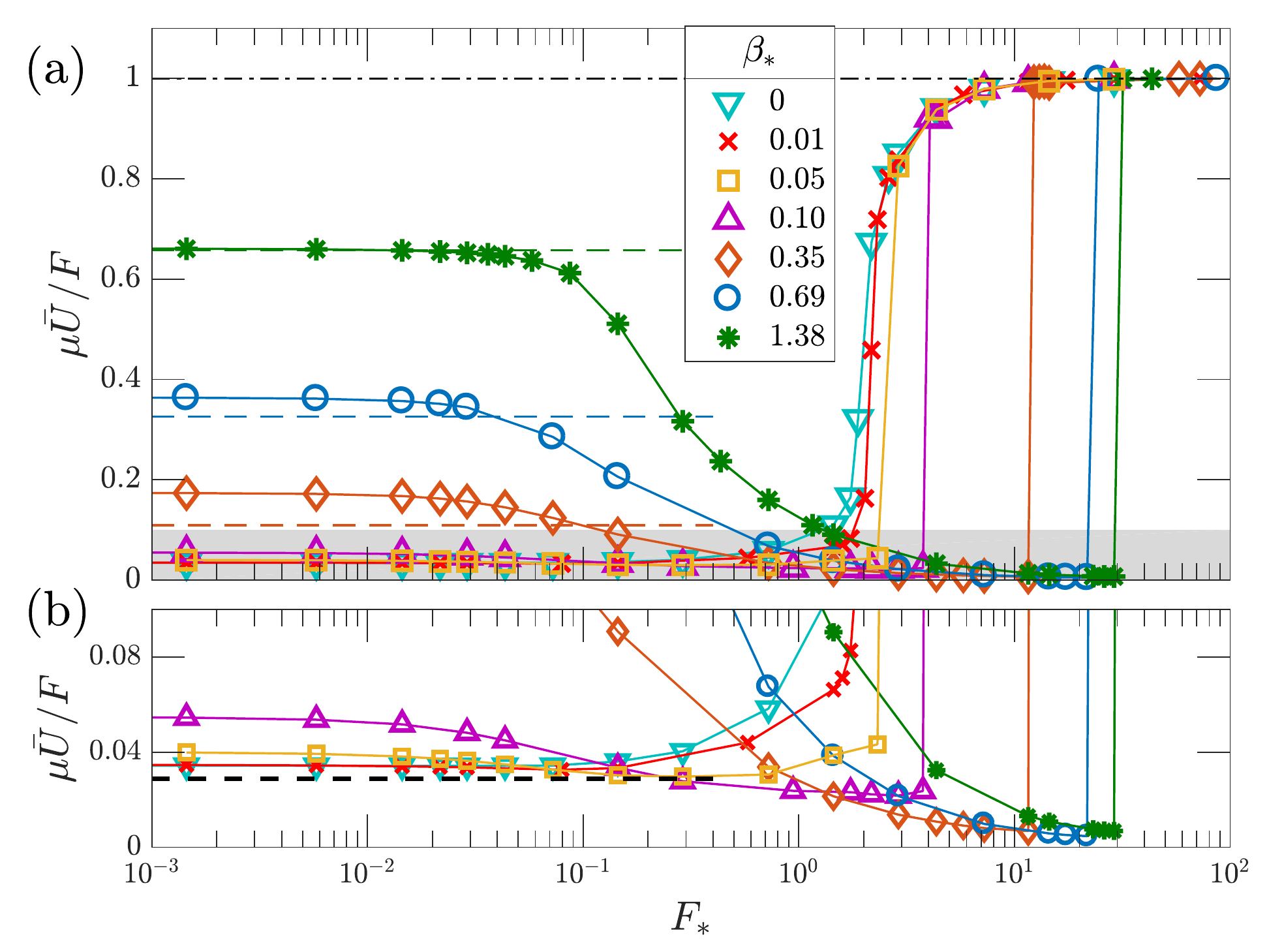}
\caption{Panel~(a) shows the ratio $\mu\bar{U}/F$ as a function of the non-dimensional forcing, $\Fs$, for seven values of $\bs$. The dashed lines are asymptotic results in~\eqref{mueff1}. Panel~(b) is a detailed view of the shaded lower part of panel~(a), showing  the eddy~saturation regime and the drag crisis. The dashed line  is the asymptotic result in~\eqref{eq:alpha_sca}.}
\label{fig:muUm_F}
%
%
\centering
\includegraphics[width = .90\textwidth]{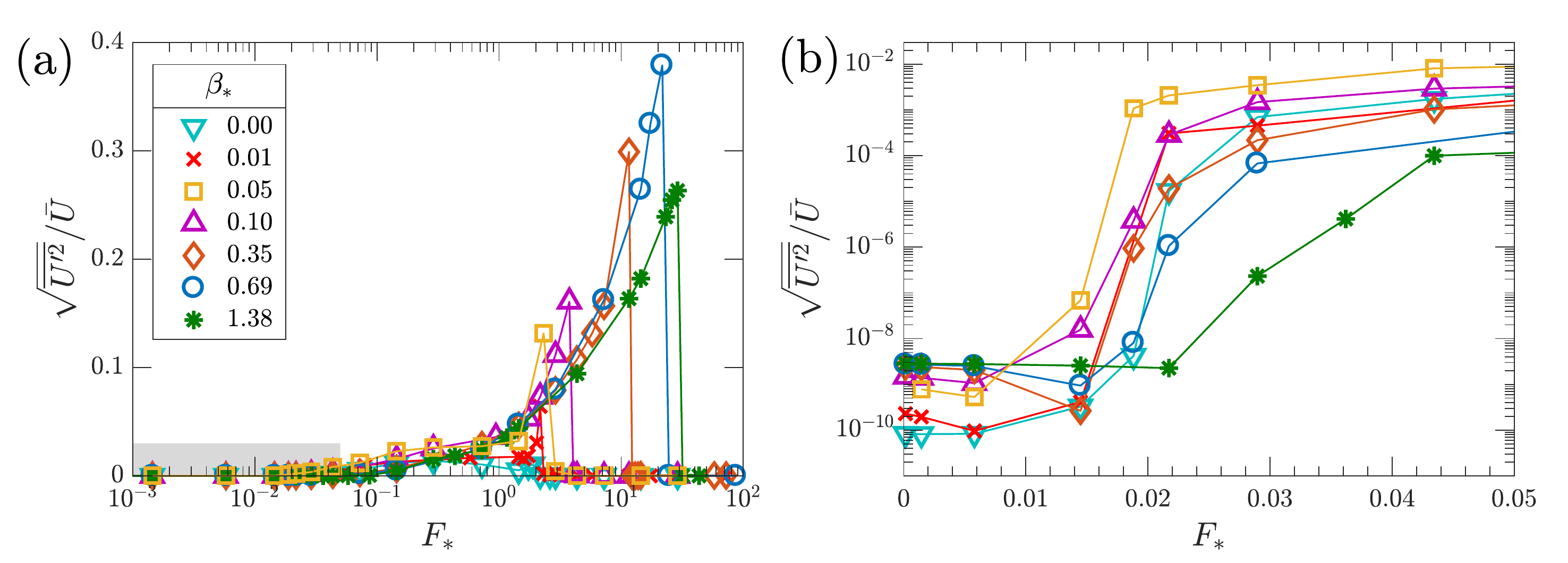}
\caption{(a) The index in~\eqref{usindex} measures the strength of the transient eddies as a function of the forcing $\Fs$. Panel (b) is  a detailed view of the shaded lower-left part of panel~(a). The onset   of transient eddies  is signaled by the large jump in the fluctuation index.}
\label{fig:dUm_Um}
%
%
\centering
\includegraphics[width = .90\textwidth]{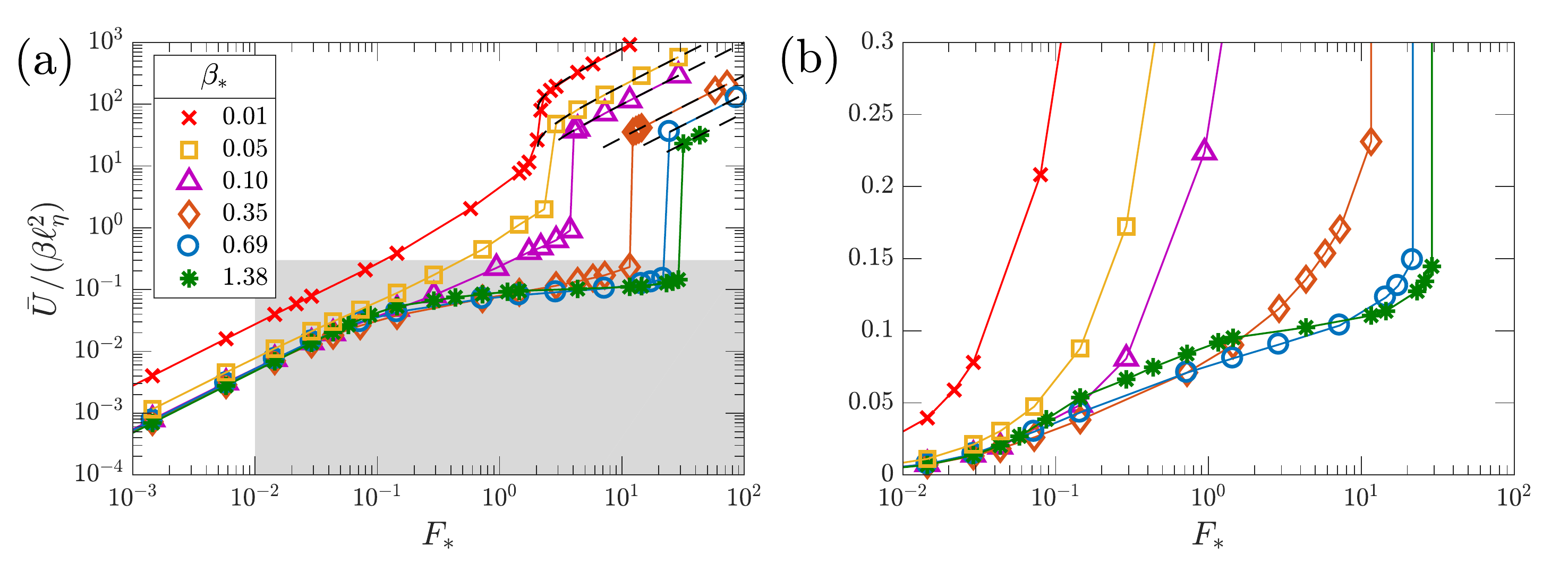}
\caption{(a)~The equilibrated large-scale flow, $\bar{U}$, scaled with $\beta\leta^2$ as a function of the non-dimensional forcing  for various values of $\bs$. Dashed curves indicate upper branch analytic result from section~\ref{sec:strong} (also scaled with $\beta\leta^2$). (b)~An expanded view of the shaded part of panel~(a) that shows the eddy~saturation regime.}\label{fig:Um_F}
\end{figure}

In this section we present a suite of solutions, all  with $\mus = 10^{-2}$. The main conclusion from these extensive calculations is that the behavior illustrated in figure~\ref{fig:U_F_b0_b20} is representative of a broad region of parameter space.

Figure~\ref{fig:muUm_F}(a) shows the ratio  $\mu\bar{U}/\wind$ as a function of $\Fs$ for seven  different values of $\beta$. The three series with $\bs\le 0.10$ are ``small-$\beta$'' cases in which closed  geostrophic contours fill most of the domain;   the other four series, with $\bs \geq 0.35$, are ``large-$\beta$'' cases in which open geostrophic contours fill most of the domain.
 For small values of $\Fs$ in figure~\ref{fig:muUm_F}(a) the flow is steady ($\psi'=U'=0$) and $\mu\bar{U}/F$ does not change with $F$:  this is the lower-branch relation~\eqref{lowerBranch} in which $\bar{U}$ varies linearly with $F$ with an effective drag coefficient $\mueff$. As $\Fs$ is increased, this steady flow becomes unstable and the strength of the transient eddy field increases with $\wind$.

Figure~\ref{fig:muUm_F}(b) shows a detailed view of the eddy saturation regime and the drag crisis. The dashed lines in the left of figures~\ref{fig:muUm_F}(a) and~(b) show the analytic results derived in section~\ref{sec:weak}. For the large-$\beta$ cases the form stress makes a very large contribution to the large-scale momentum balance prior the drag crisis.  
We emphasize that although drag $\mu$ does not directly balance $\wind$ in this regime, it does play a crucial role in producing non-zero form stress $\fsb$. In all of the solutions summarized in figure~\ref{fig:muUm_F}, non-zero $\mu$ is required so that the flow is asymmetric upstream and downstream of topographic features; this asymmetry induces non-zero $\fsb$.

In figure~\ref{fig:dUm_Um} we use
\beq
 {\sqrt{\overline{U'^2}}}\Big\slash{\bar U} \label{usindex}
\eeq
as an indication of the onset of the transient-eddy instability and as an index of the strength of the transient eddies. Remarkably, the onset of the instability is roughly at $\Fs= 1.5\times10^{-2}$ for all values of $\bs$: the onset of transient eddies is the sudden  increase in~\eqref{usindex} by a factor of about $10^4$ or $10^5$ in figure~\ref{fig:dUm_Um}(b). The transient eddies result in reduction of $\mu\bar{U}/F$; for the large-$\beta$ runs, this is the eddy~saturation regime. In the presentation in figure~\ref{fig:muUm_F}(a) the eddy~saturation regime is the decease in $\mu\bar{U}/\wind$ that occurs once $0.03< \Fs<0.3$ (depending on $\bs$). The eddy~saturation regime is terminated by the drag-crisis jump to the upper branch where $\mu \bar U/F \approx 1$. This coincides with vanishing of the transients: on the upper branch the flow becomes is steady: $\psi'=U'=0$: see~figure~\ref{fig:dUm_Um}(a).


\begin{figure}
	\centering
	\includegraphics[width = 0.9\textwidth,trim={0cm 0cm 0cm 0mm},clip]{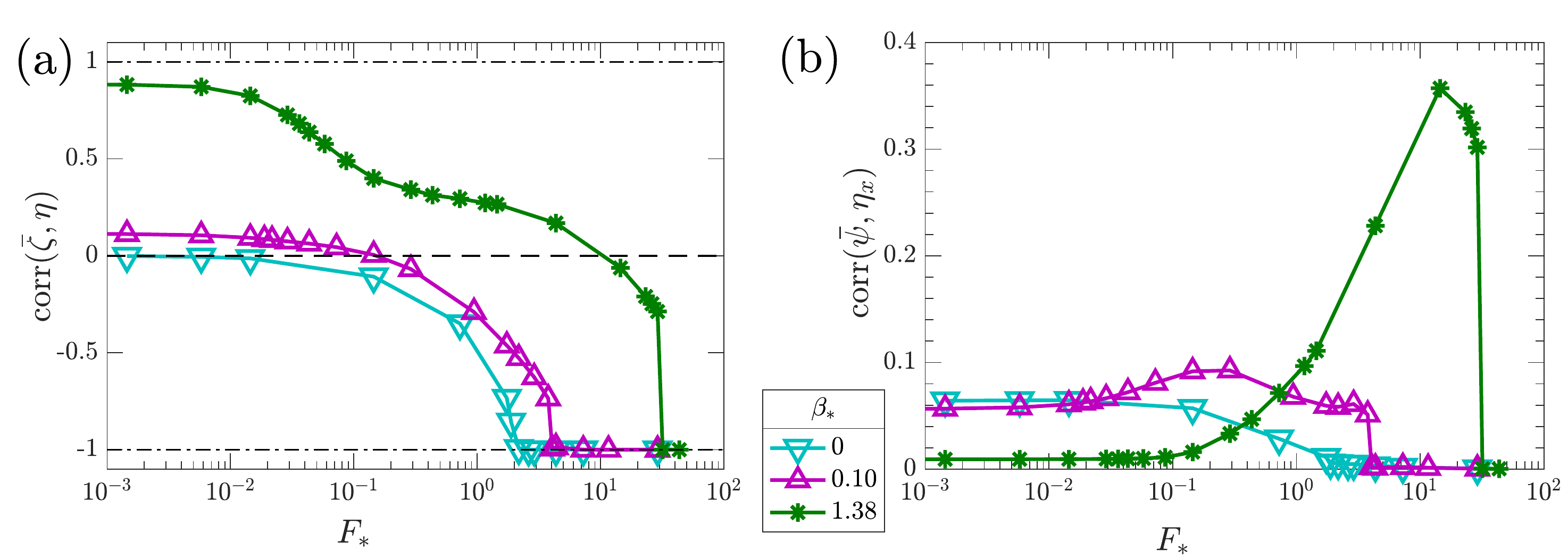}
	\vspace{-0.5em}
	\caption{(a) Correlation of the standing eddy vorticity $\bar \zeta$  with the topographic PV $\h$ for $\bs=0$,~$0.10$ and $1.38$. For $\bs=0$ the correlation $\corr(\bar\zeta,\h)$ is always negative; for $\Fs=10^{-3}$, $\corr(\bar\zeta,\eta)=-1.35\times10^{-3}$. (b) Correlation of  $\bar\psi$ with $\eta_x$.}\label{fig:mean_corrs}
\end{figure}

Figure~\ref{fig:Um_F} shows the eddy~saturation regime that is characteristic of the three series with $\bs \geq 0.35$. Eddy~saturation occurs for forcing in the range
$$
 0.1 \lessapprox \Fs \lessapprox 30\, ;
$$
in this regime the large-scale flow is limited to the relatively small range
$$
 0.06\, \beta\leta^2\lessapprox\bar U \lessapprox 0.25\, \beta\leta^2 \per
$$
In anticipation of analytic results from the next section we note that in the relation above, $\beta\leta^2$ is the speed of Rossby waves excited by this topography with typical length scale $\leta$.

Figure~\ref{fig:mean_corrs} shows the correlations $\corr(\bar\zeta,\h)$ and $\corr(\bar\psi,\eta_x)$ as a function of the forcing $\Fs$ for three values of $\bs$. In most weakly forced cases $\bar\zeta$ is positively correlated with $\h$; as the forcing $F$ increases, $\bar \zeta$ and $\h$ become anti-correlated. However, for $\bs=0$ the correlation $\corr(\bar\zeta,\eta)$ is negative for \emph{all} values of $F$: for the monoscale topography used here, the term~$\la\eta\D\bar\zeta\ra$, which is the only source of enstrophy in the time-average of~\eqref{eq:dQdt} if $\beta=0$, can be approximated as $(\mu+\nu/\leta^2) \la\bar\zeta\h\ra$. Therefore in this case $\la\bar\zeta\h\ra$ must be negative (see the discussion in Appendix~\ref{app:balances}).

\section{A quasilinear (QL)  theory\label{sec:QL}}

A prediction of the statistical steady state of~\eqref{eq:U_t} and~\eqref{eq:z_NL} was first made by \citet{Davey-1980a}. In this section we present Davey's quasilinear (\quasilinear{}) theory and in subsequent sections we  explore its validity in various regimes documented in section~\ref{survey}.  QL is an exploratory  approximation  obtained  by retention of all the terms consistent with easy analytic solution of the  QGPV equation: see \eqref{eq:z_QL} below;  terms  hindering analytic solution are  discarded without \textit{a priori} justification. We show in sections~\ref{sec:weak} and~\ref{sec:strong} that \quasilinear{} is in good agreement with numerical solutions in some parameter ranges e.g., everywhere on the upper branch and on the lower branch provided that $\bs \gtrapprox 1$. With hindsight, and by comparison with the numerical solution, one can understand these QL successes~\textit{a posteriori} by showing that the terms discarded to reach~\eqref{eq:z_QL} are, in fact, small relative to at least some of the retained terms.

\begin{figure}
\centering
\includegraphics[width = 0.85\textwidth,trim={0cm 0cm 0cm 7mm},clip]{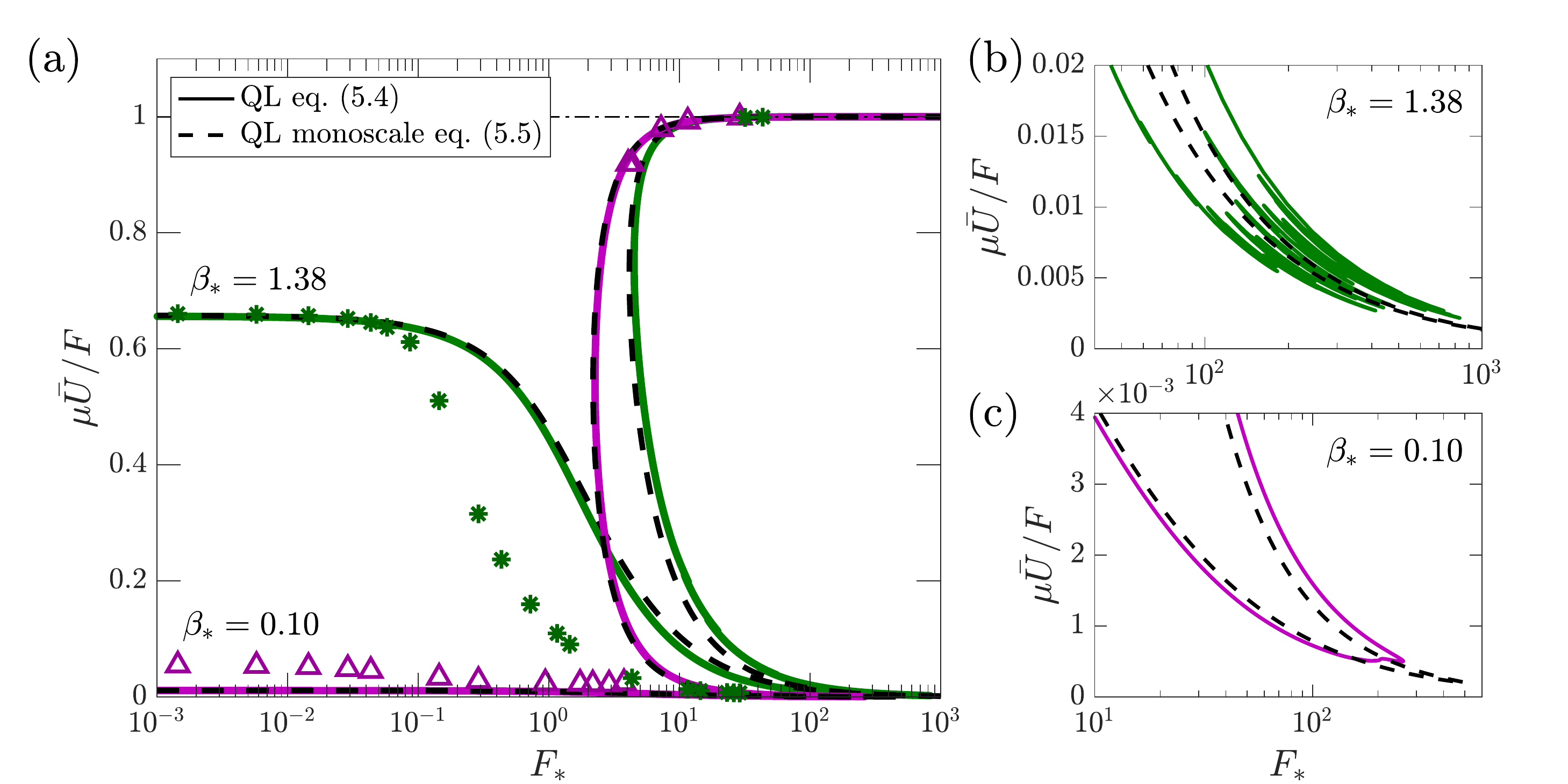}
\vspace{-0.5em}\caption{(a) The large-scale flow, $\mu\bar{U}/F$, as a function of forcing the $\Fs$ for the cases with $\bs=0.10$~and~1.38. The solid curves are the \quasilinear{} predictions using a single realization to evaluate the sum in~\eqref{eq:sigma_QLf} and the dashed curves are the ensemble-average predictions from~\eqref{eq:sigma_QLf_mono};  the markers indicate the numerical solution of the full nonlinear system~\eqref{eq:U_t} and~\eqref{eq:z_NL}. Panels~(b) and (c) show a detailed view of the bottom right corner of panel (a);  the resonances in the denominator of~\eqref{eq:sigma_QLf} come into play in this small region.}\label{fig:QL_fig}
\end{figure}

Assume that the QGPV equation~\eqref{eq:z_NL} has  a steady solution and also neglect  $\J(\psi,q)=\J(\psi,\eta)+\J(\psi,\zeta)$. These \textit{ad hoc} approximations result in the \quasilinear{}  equation
\beq
 U\zeta_x + \beta\psi_x +\mu\zeta = - U\h_x\com\label{eq:z_QL}
\eeq
in which $U$ is determined by the steady mean flow equation
\beq
 F-\mu U- \fs=0\per\label{eq:U_NL_static}
\eeq
In~\eqref{eq:z_QL} we have neglected lateral dissipation so that the dissipation is $\D=\mu$ (see discussion in section~\ref{sec:form}). Notice that the only nonlinear term in~\eqref{eq:z_QL} is $U\zeta_x$. Regarding $U$ as an unknown parameter, the solution of~\eqref{eq:z_QL} is:
\beq
 \psi =U \sum_{\bk}\frac{\ii k_x \h_{\bk}\ee^{\ii \bk \bcdot \bx}}{\mu |\bk|^2 - \ii k_x (\beta-|\bk|^2U)} \per\label{eq:psi_QLf}
\eeq
Thus the \quasilinear{} approximation to the form stress in~\eqref{eq:U_NL_static}  is
\beq
 \la\psi\eta_x\ra = U \sum_{\bk}\frac{ \mu  k_x^2|\bk|^2 |\h_{\bk}|^2}{\mu^2 |\bk|^4 + k_x^2 (\beta-|\bk|^2U)^2}\per\label{eq:sigma_QLf}
\eeq
Inserting~\eqref{eq:sigma_QLf} into the large-scale momentum equation~\eqref{eq:U_NL_static} one obtains an equation for $U$. This equation is a polynomial of order $2N+1$, where $N \gg 1$ is the number of non-zero terms in the sum in~\eqref{eq:sigma_QLf}. This implies, at least in principle, that there might be  many real solutions for $U$. However, for the monoscale topography of figure~\ref{fig:topo}, we usually find either one real solution or three as $F$ is varied: see figure~\ref{fig:QL_fig}(a). Only in a very limited parameter region we find a multitude of additional real solutions: see figure~\ref{fig:QL_fig}(b). The fine-scale features evident in figure~\ref{fig:QL_fig}(b) vary greatly between different realizations of the topography and are irrelevant for the full nonlinear system.

For the special case of isotropic monoscale topography we simplify~\eqref{eq:sigma_QLf} by converting the sum over $\bk$ into an integral that can be evaluated analytically (see appendix~\ref{app:fs_mono}). The result is
\beq
\la\psi\eta_x\ra =  \frac{\mu U \leta^2\etarms^2}{\mu^2\leta^2 +\left(\beta\leta^2 -U\right)^2 +\mu\leta \sqrt{\mu^2\leta^2 +\left(\beta\leta^2 -U\right)^2}}\per
\label{eq:sigma_QLf_mono}
\eeq
Expression~\eqref{eq:sigma_QLf_mono} is a good approximation to the sum~\eqref{eq:sigma_QLf} for the monoscale topography of figure~\ref{fig:topo}(a) that has power over an annular region in wavenumber space with width $\Delta k\,L\approx 8$. The dashed curves in figure~\ref{fig:QL_fig} are obtained by solving the mean-flow equation~\eqref{eq:U_NL_static} with form stress given by the analytic expression in~\eqref{eq:sigma_QLf_mono}; there is good agreement with the sum~\eqref{eq:sigma_QLf} except in the small regions shown in panels~(b) and~(c), where the resonances of the denominator come into play. This comparison shows that the form stress produced in a single realization of random topography is self-averaging i.e., the ensemble average in~\eqref{eq:sigma_QLf_mono} is close to the result obtained by evaluating the sum in~\eqref{eq:sigma_QLf} using a single realization of the $\eta_{\bk}$'s.

Figure~\ref{fig:QL_fig} also compares the \quasilinear{} prediction in~\eqref{eq:sigma_QLf} and~\eqref{eq:sigma_QLf_mono} to solutions of the full system. Regarding weak forcing ($\Fs\ll 1$), the \quasilinear{} approximation seriously  underestimates $\mu U/F$ for the case with $\bs=0.1$ in figure~\ref{fig:QL_fig}. The failure of the \quasilinear{} approximation in this case with dominantly closed geostrophic contours is expected because the important  term $\J(\psi,\h)$ is discarded in~\eqref{eq:z_QL}. On the other hand, the \quasilinear{} approximation has some success for the case with $\bs=1.38$: proceeding in figure~\ref{fig:QL_fig} from very small $\Fs$, we find close agreement till about $\Fs \approx 0.1$. At that point the \quasilinear{} approximation departs from the full solution: the velocity $U$ predicted by the \quasilinear{} approximation is greater than the actual velocity, meaning that the \quasilinear{} form stress $\fsb$ is too small. This failure of the \quasilinear{} approximation is clearly associated with the linear instability of the steady solution and the development of transient eddies: the nonlinear results for the $\bs=1.38$ case in figure~\ref{fig:QL_fig}(a) first depart from the \quasilinear{} approximation when the index~\eqref{usindex} signals the onset of unsteady flow.  This failure of the \quasilinear{} theory due to transient eddies will be further discussed in section~\ref{sec:saturation}. For strong forcing ($\Fs\gg1$), the \quasilinear{} approximation predicts very well the upper branch solution.

The heuristic  assumptions leading to the \quasilinear{} estimate~\eqref{eq:sigma_QLf} are drastic. But we will see in sections~\ref{sec:weak} and~\ref{sec:strong}, the \quasilinear{} approximation captures the qualitative behavior of the full numerical solution and, in some parameter regimes such as $\bs \gtrapprox1$, even provides a good quantitative prediction of $\bar U$.

\section{The weakly forced regime, $\Fs \ll 1$ \label{sec:weak}}

In this section we consider the weakly forced case. In figures~\ref{fig:U_F_b0_b20} and~\ref{fig:muUm_F} this regime is characterized by the ``effective drag" $\mueff$ in~\eqref{lowerBranch}. Our main goal here is to determine $\mueff$ in the weakly forced regime.

Reducing the strength of the forcing $\Fs$ to zero is equivalent to taking a limit in which the system is linear. This weakly forced flow is then  steady, $\psi'=0$, and terms which are quadratic in the flow fields $U$ and $\psi$, namely $U \zeta_x$ and $J(\psi,\zeta)$, are negligible. Thus in the  limit $\Fs \to 0$ the eddy field satisfies the steady linearized QGPV equation:
\begin{align}
	 \J\(\psi,\eta\) + \beta\psi_x +\mu \zeta &=-U\eta_x \per\label{eq:z_NL_smallF}
\end{align}
When compared to the \quasilinear{} approximation~\eqref{eq:z_QL} we see that~\eqref{eq:z_NL_smallF} contains  the additional linear term $\J\(\psi,\eta\)$ and does not contain the non-linear term $U\zeta_x$.  We regard the right hand side of the linear equation~\eqref{eq:z_NL_smallF} as  forcing that generates the streamfunction $\psi$.

\subsection{The case with either $\mus\gg 1$ or $\bs\gg1$\label{sec:weak_QL}}

\begin{figure}
\centering
\includegraphics[width = .5\textwidth,trim={0cm 0cm 0cm 7mm},clip]{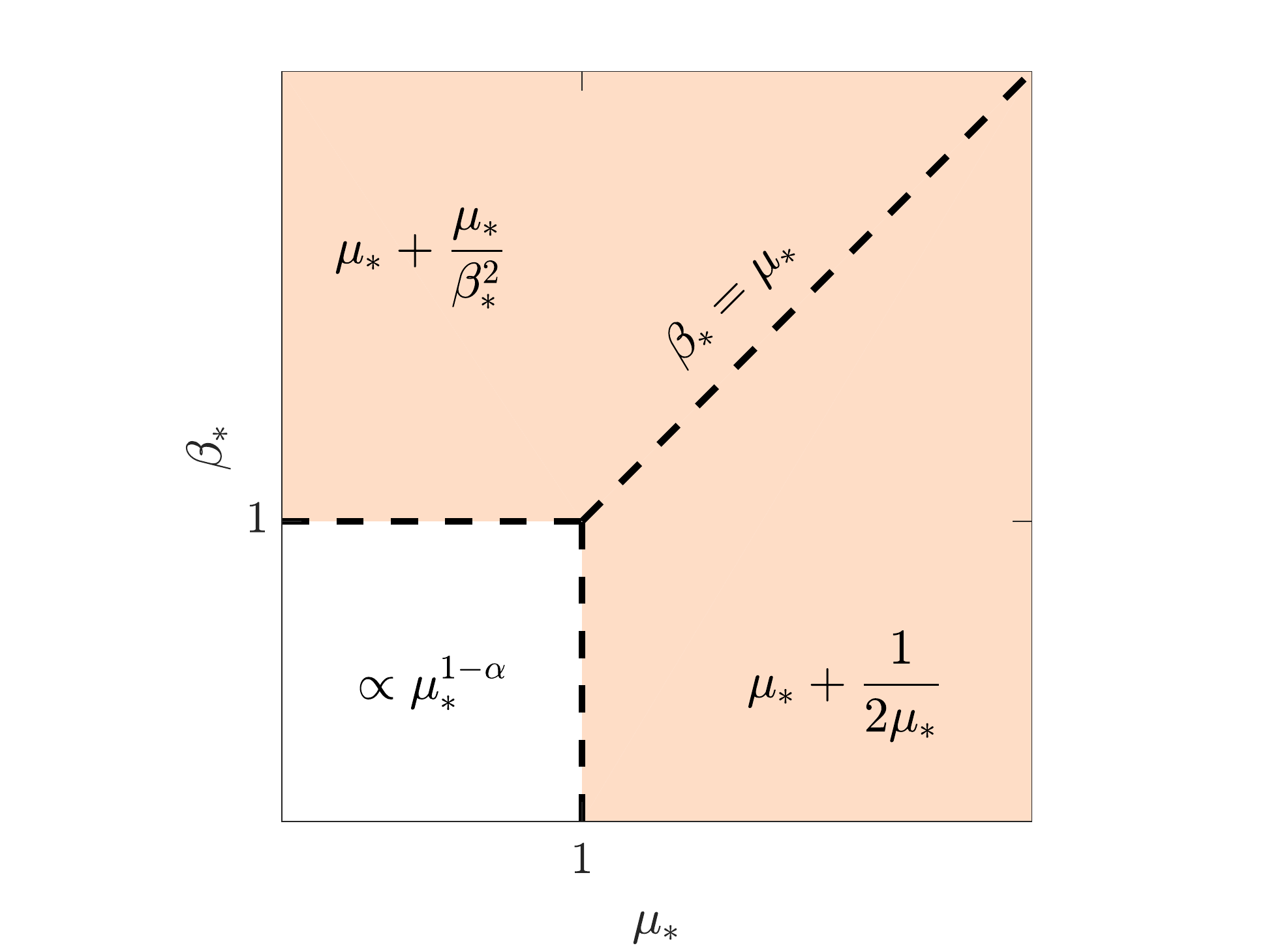}
\vspace{-1em}
\caption{ Schematic for the three different parameter regions in the weakly forced regime. Shaded region depicts the parameter range for which the \quasilinear{} theory gives good predictions. For $\bs<1$ and $\mus<1$ the form stress and the large-scale flow largely depend on the actual geometry of the topography and $J(\psi,\eta)$ cannot be neglected. The expressions show the behavior of ${\mueff}_*=\mueff/\etarms$ in each parameter region.}\label{fig:weak_schem}
\end{figure}

Assuming that lengths scale with $\leta$, the ratio of the terms on the left of~\eqref{eq:z_NL_smallF} is:
\beq
\beta\psi_x\Big/\J\(\psi,\eta\) = O(\bs)  \qquad \text{and} \qquad \mu \zeta\Big/\J\(\psi,\eta\)= O(\mus)\per
\eeq
If $\mus\gg 1$ or if $\bs\gg1$ then  $\J\(\psi,\eta\)$  is negligible relative to one, or both, of the other two terms on the left hand side of~\eqref{eq:z_NL_smallF}. In that case,  one can neglect the Jacobian in~\eqref{eq:z_NL_smallF} and   adapt the \quasilinear{} expression~\eqref{eq:sigma_QLf_mono} to determine the effective drag of monoscale topography as
 \beq
\mueff = \mu +  \frac{\mu  \etarms^2  \leta^2 }{\mu^2\leta^2 +\beta^2\leta^4+\mu \leta \sqrt{\mu^2\leta^2+\beta^2\leta^4 }}\per
 \label{mueff1}
 \eeq
In simplifying the \quasilinear{} expression~\eqref{eq:sigma_QLf_mono} to the linear result~\eqref{mueff1} we have neglected $U$ relative to either $\beta \leta^2$ or $\mu \leta$: this simplification is appropriate in the limit $\Fs \to 0$. The expression in~\eqref{mueff1} is accurate within the shaded region in figure~\ref{fig:weak_schem}. The dashed lines in figures~\ref{fig:U_F_b0_b20} and~\ref{fig:muUm_F}(a) that correspond to the series with $\bs\ge0.35$ indicate the approximation $\bar U \approx \wind/\mueff$ with $\mueff$ in~\eqref{mueff1}.

\subsection{The thermal analogy --- the case with $\mus\lessapprox 1$ and $\bs\lessapprox 1$}

When both $\mus$ and $\bs$ are order one or less the term $\J(\psi,\eta)$ in~\eqref{eq:z_NL_smallF} cannot be neglected. As a result of this Jacobian, the weakly forced regime cannot be recovered as a special case of the \quasilinear{} approximation. In this interesting case we rewrite~\eqref{eq:z_NL_smallF} as
\beq
\J\(\eta + \beta y, \psi-U y\) = \mu \nabla^2 \psi \com
\label{Welander1}
\eeq
and rely on intuition based on  the ``thermal analogy". To apply the analogy we regard $\eta + \beta y$ as an effective steady streamfunction advecting a passive scalar $\psi - U y$. The planetary vorticity gradient $\beta$ is analogous to a large-scale zonal flow $-\beta$ and the large-scale flow $U$ is analogous to a large-scale tracer gradient; the drag $\mu$ is equivalent to the diffusivity of the scalar. The form stress $\fs$ is analogous to the meridional flux of tracer $\psi$ by the meridional velocity $\eta_x$. Usually in the passive-scalar problem the large-scale tracer gradient $U$ is imposed and the main goal is to determine the flux $\fs$ (equivalently the Nusselt number). But here, $U$ is unknown and must be determined by satisfying the steady version of the large-scale momentum equation~\eqref{eq:U_NL_static}. Geostrophic contours are equivalent to streamlines in the thermal analogy.

\begin{figure}
\centering
\includegraphics[width = .95\textwidth,trim={0cm 1cm 0cm 0},clip]{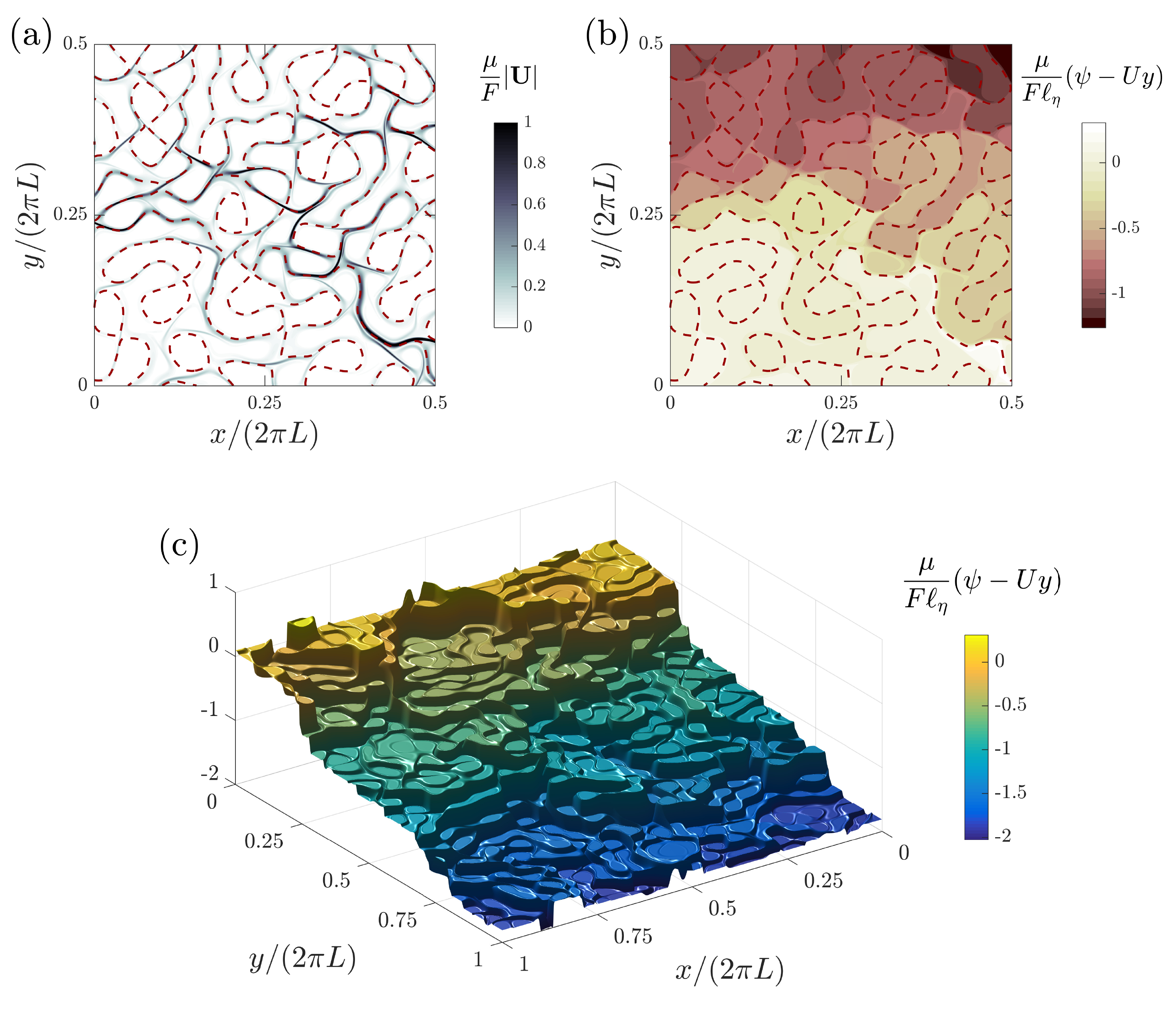}
\caption{Snapshots of flow fields for weakly forced simulations at $\Fs=10^{-3}$ with dissipation $\mus=5\times10^{-3}$ and $\bs=0$. (a)~The total velocity magnitude $|\bU|$. The flow is restricted to a boundary layer around the dashed $\h=0$ contour. (b)~The total streamfunction $\psi-Uy$ for the solution in panel (a). (c)~Surface plot of the total streamfunction, $\psi-Uy$. The terraced hillside structure is apparent. (In panels~(a) and~(b) only one quarter of the domain is shown.)}
\label{fig:Nu_mono}
%
\centering
\includegraphics[width = .6\textwidth]{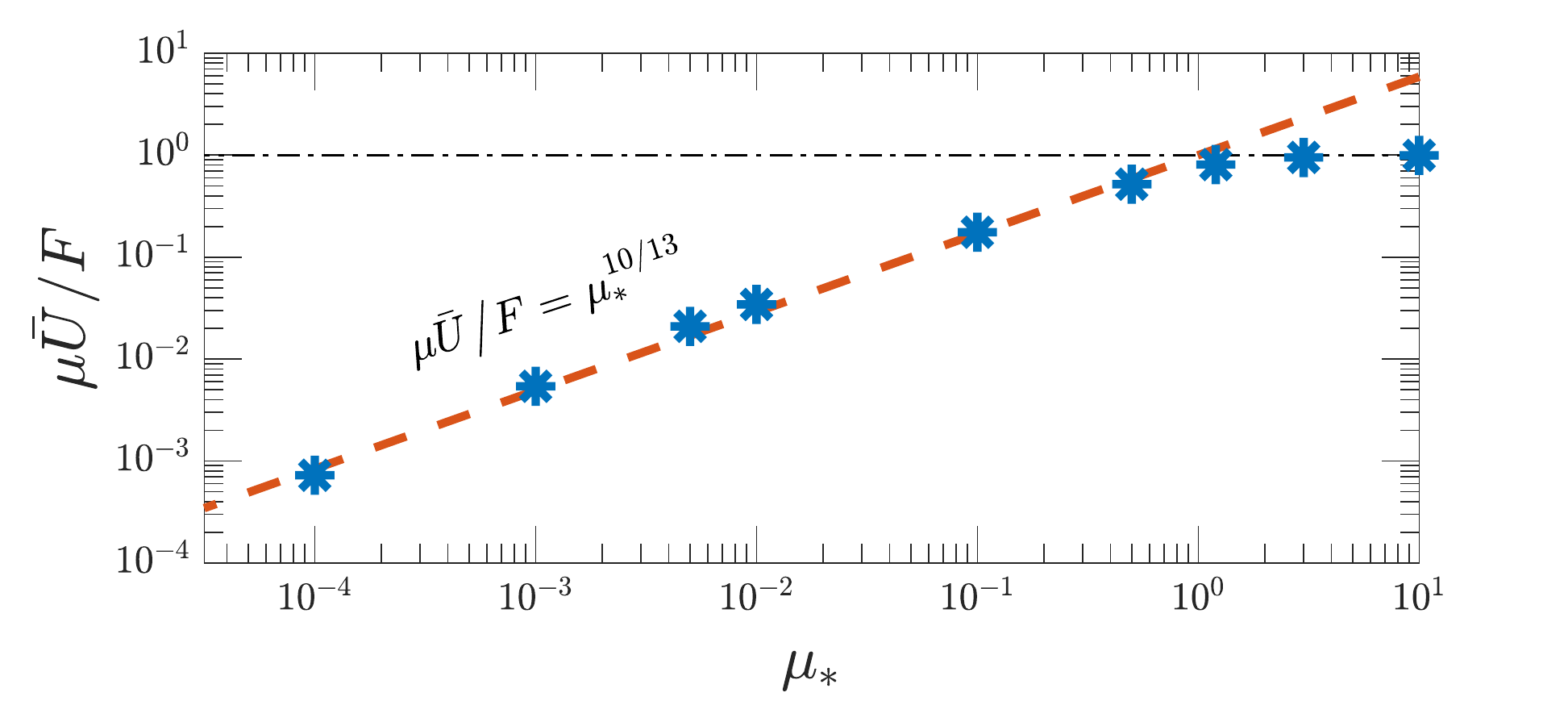}
\vspace{-0.7em}\caption{The large-scale flow for weakly forced solutions ($\Fs=10^{-3}$) with $\beta=0$ as a function of $\mus$. The dashed line shows the scaling law~\eqref{eq:alpha_sca} with $c=1$.}
\label{fig:Nu_b0_sca}
\end{figure}

With the thermal analogy, we can import results from the passive-scalar problem. For example, in the passive-scalar problem, at large  P\'eclet number, the scalar is uniform within closed streamlines  \citep{Batchelor-1956, Rhines-Young-1983}. The analog of this ``Prandtl--Batchelor theorem'' is that in the limit $\mus \to 0$  the total streamfunction, $\psi -Uy$, is constant within \emph{any} closed geostrophic contour i.e., all parts of the domain contained within closed geostrophic contours are stagnant; see also the paper by  \citet{Ingersoll-1969}. This ``Prandtl--Batchelor theorem'' explains the result in figure~\ref{fig:Nu_mono}, which shows a weakly forced, small-drag solution with $\bs=0$. The domain is packed with stagnant eddies (constant $\psi -Uy$) separated by thin boundary layers. The ``terraced hillside'' in figure~\ref{fig:Nu_mono}(c) is even more striking than in figure~\ref{fig:eg_closed_B}: the solution in figure~\ref{fig:eg_closed_B} has transient eddies resulting a blurring of the terraced structure. The weakly-forced solution in figure~\ref{fig:Nu_mono} is steady and the thickness of the steps between the terraces is limited only by the small drag, $\mus = 5\times10^{-3}$.

\citet{Isichenko-etal-1989} and \citet{Gruzinov-etal-1990} discuss the effective diffusivity of a passive scalar due to advection by a steady monoscale streamfunction. Using a scaling argument, \citet{Isichenko-etal-1989} show that in the high P\'eclet number limit the effective diffusivity of a steady monoscale flow  is $D_{\mathrm{eff}} = D P^{10/13}$, where $D$ is the small  molecular diffusivity and $P$ is the P\'eclet number; the exponent $10/13$ relies on critical exponents determined by percolation theory. Applying Isichenko's  passive-scalar results  to the $\beta=0$ form-stress problem  we obtain the scaling
 \beq
\mueff = c \mu^{3/13} \etarms^{10/13} \com \qquad \text{and} \qquad \frac{\mu \bar{U}}{F}=  \frac{1}{c} \(\frac{\mu}{\etarms}\)^{10/13}\com \label{eq:alpha_sca}
 \eeq
where $c$ is a dimensionless constant. Numerical solutions of~\eqref{eq:U_t} and~\eqref{eq:z_NL} summarized in figure~\ref{fig:Nu_b0_sca} confirm this remarkable ``ten-thirteenths" scaling and show that the constant $c$ in~\eqref{eq:alpha_sca} is close to one. The dashed lines in figures~\ref{fig:U_F_b0_b20} and~\ref{fig:muUm_F}(b),  corresponding to the solution suites with $\bs\leq 0.1$, show the scaling law~\eqref{eq:alpha_sca} with $c=1$.

To summarize: the weakly forced regime is divided into the easy large-$\beta$ case, in which $\mueff$ in~\eqref{mueff1} applies, and the more  difficult case with small or zero $\beta$. In the difficult case, with closed geostrophic contours, the thermal analogy and the Prandtl-Batchelor theorem  show that the flow is partitioned into stagnant dead zones; Isichenko's $\beta=0$ scaling law in~\eqref{eq:alpha_sca} is the main results in this case. The value of $\bs$ separating these two regimes in the schematic of figure~\ref{fig:weak_schem} is identified with the  $\beta$ below which~\eqref{mueff1} underestimates $\mu U/F$ compared to~\eqref{eq:alpha_sca}. For the topography used in this work, and taking $c=1$ in~\eqref{eq:alpha_sca}, this is $\bs=0.17$. (If we choose $c=0.5$ the critical value is $\bs=0.24$.) This rationalizes why the $\beta=0$ result in~\eqref{eq:alpha_sca} works better than $\mueff$ in~\eqref{mueff1} for $\bs<0.35$: see figure~\ref{fig:muUm_F}.

\section{The strongly forced regime, $\Fs \gg 1$ \label{sec:strong}}

We turn now to the upper  branch, i.e., to  the flow beyond the drag crisis. In this strongly forced regime the flow is steady: $\psi'=U'=0$ and the \quasilinear{} theory gives good results for all values of $\bs$.

\begin{figure}
\centering
\includegraphics[width = .65\textwidth]{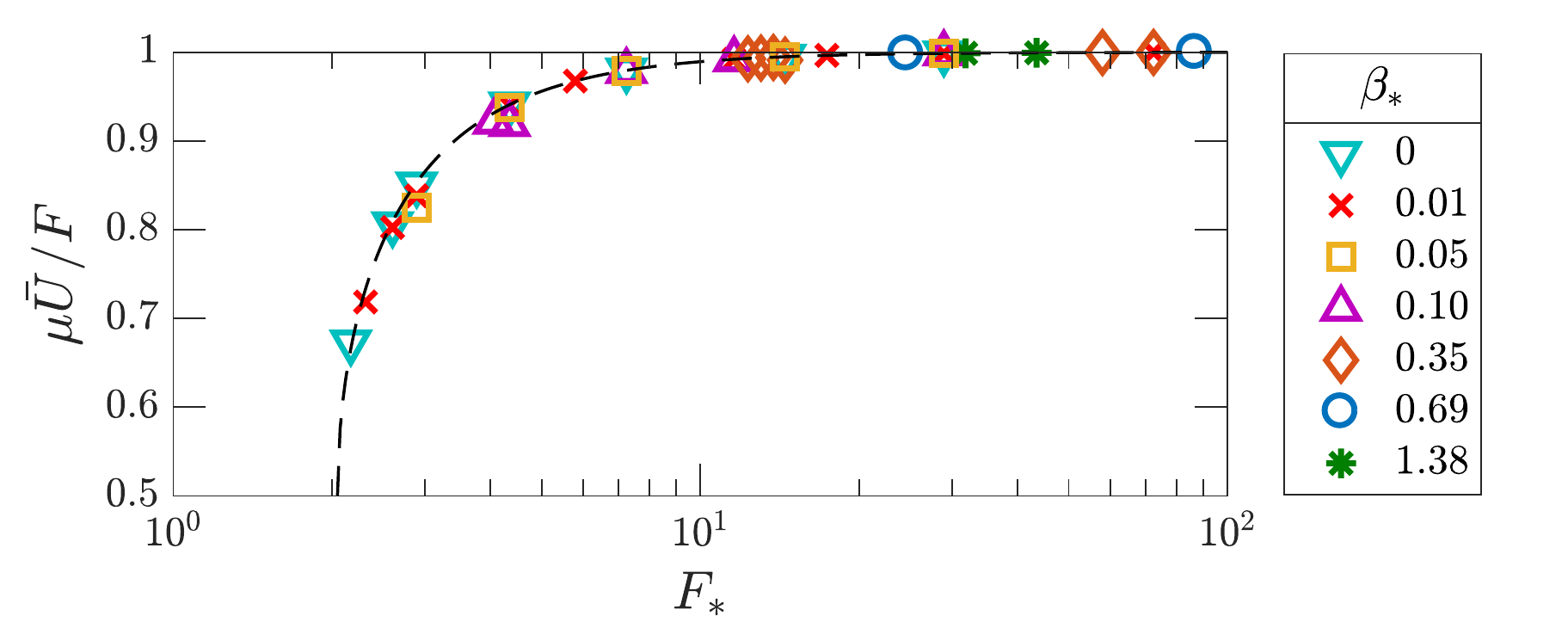}
\vspace{-.7em}\caption{A detailed view of the upper-branch flow regime i.e., the upper-right part of figure~\ref{fig:muUm_F}(a), together with the analytic prediction~\eqref{eq:U_largeF} (dashed).}\label{fig:upper_branch}
\end{figure}

The  solutions in figure~\ref{fig:Um_F}(a) show that on the upper branch the large-scale flow $\bar U$ is much faster than the phase speed of Rossby waves excited by the topography, i.e., $\bar U\gg \beta\leta^2$. Therefore, we can simplify the QL approximation in~\eqref{eq:sigma_QLf_mono} by neglecting terms smaller than $\beta\leta^2/U$. This gives:
\begin{align}
\la\psi\eta_x\ra &=\frac{\mu\etarms^2\leta^2}{U} +O(\beta\leta^2/U)^2\per\label{eq:fs_largeU_mono}
\end{align}
This result is independent of $\beta$ up to $O(\beta\leta^2/U)^2$. Using~\eqref{eq:fs_largeU_mono} in the large-scale zonal momentum equation~\eqref{eq:U_NL_static}, while keeping in mind that $0\le U\le F/\mu$, we solve a quadratic equation for $U$ to obtain:
\beq
\frac{\mu U}{F}=\frac{1}{2} + \sqrt{\frac1{4}-\frac1{\Fs^2}}\per\label{eq:U_largeF}
\eeq
The location of the drag crisis depends on $\beta$, and on details of the topography that are beyond the reach of the \quasilinear{} approximation. But once the solution is on the upper branch these complications are irrelevant e.g.,~\eqref{eq:U_largeF} does not contain $\beta$. The dashed curve in figure~\ref{fig:upper_branch} compares~\eqref{eq:U_largeF} to numerical solutions of the full system and shows close agreement.

We get further intuition about the structure of the upper-branch flow through the \quasilinear{} equation~\eqref{eq:z_QL}. For large $U$ we have a two-term balance in~\eqref{eq:z_QL} that gives $\bar\zeta\approx-\h$, so that $\bar q$ is $O(\leta\etarms^{-2}U^{-1})$. Figure~\ref{fig:mean_corrs}(a) shows that on the upper branch the  correlation of $\bar \zeta$ with $\h$ is close to $-1$ and  numerical upper-branch solutions confirm that $\bar\zeta\approx\-\eta$.

%

\section{Intermediate forcing:  eddy~saturation and the drag crisis\label{sec:saturation}}

In sections~\ref{sec:weak} and~\ref{sec:strong} we discussed  limiting cases with small and large forcing respectively. In both these limits the solution is steady i.e., there are no transient eddies. We now turn to the more complicated situation with forcing of intermediate strength. In this regime the solution has transient eddies and numerical solution shows that these produce drag that is additional to the \quasilinear{} prediction (see figure~\ref{fig:QL_fig} and related discussion). The eddy saturation regime, in which $U$ is insensitive to large changes in  $\Fs$ (see figure~\ref{fig:Um_F}), is also characterized by forcing of intermediate strength: the solution described in section~\ref{sec:example_open} is an example. Thus a goal is to better understand the eddy saturation regime and its termination by the drag crisis.

\subsection{Eddy saturation regime\label{keff_th}}

As wind stress increases transient eddies emerge: in figure~\ref{fig:dUm_Um} this instability of the steady solution occurs very roughly at $\Fs=1.5\times10^{-2}$ for all values of $\beta$. The power integrals in appendix~\ref{app:balances} show that the transient eddies gain kinetic energy from the standing eddies $\bar\psi$ through the conversion term $\la\bar{\psi}\bnabla\bcdot\bE\ra$, where
\beq
 \bE\defn \overline{\bU'q'}\com\label{eq:Evector}
\eeq
is the time-averaged eddy PV flux. The conclusions from appendix~\ref{app:balances} are summarized in figure~\ref{fig:E_Q_diagram} by showing the energy and enstrophy transfers among the four flow components $\bar U$, $U'$, $\bar\psi$ and $\psi'$.

\begin{figure}
\centering
\includegraphics[width = 0.85\textwidth]{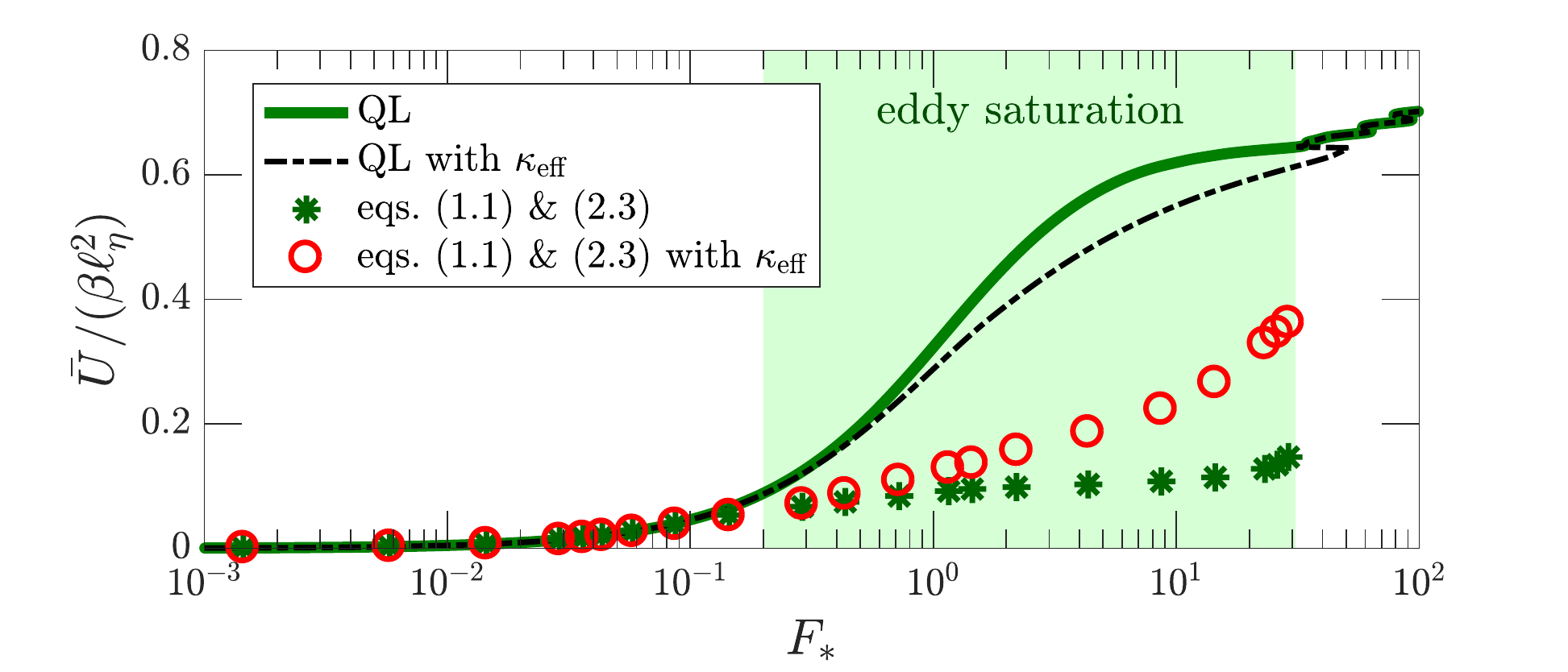}
\vspace{-1em}\caption{The eddy saturation regime for $\bs=1.38$ (shaded). Asterisks $*$ indicate numerical solutions of~\eqref{eq:U_t} and~\eqref{eq:z_NL}. Circles $\circ$ show the numerical solutions of~\eqref{eq:U_t} and~\eqref{eq:z_NL} with the added PV diffusion, $\keff\nabla^2 q$. The solid curve is the \quasilinear{}  prediction~\eqref{eq:sigma_QLf} and the dashed-dot curve is the \quasilinear{} prediction with added PV diffusion. }\label{fig:QL_b20_nueff_1p}
\vspace{0.5em}

\includegraphics[width = .85\textwidth]{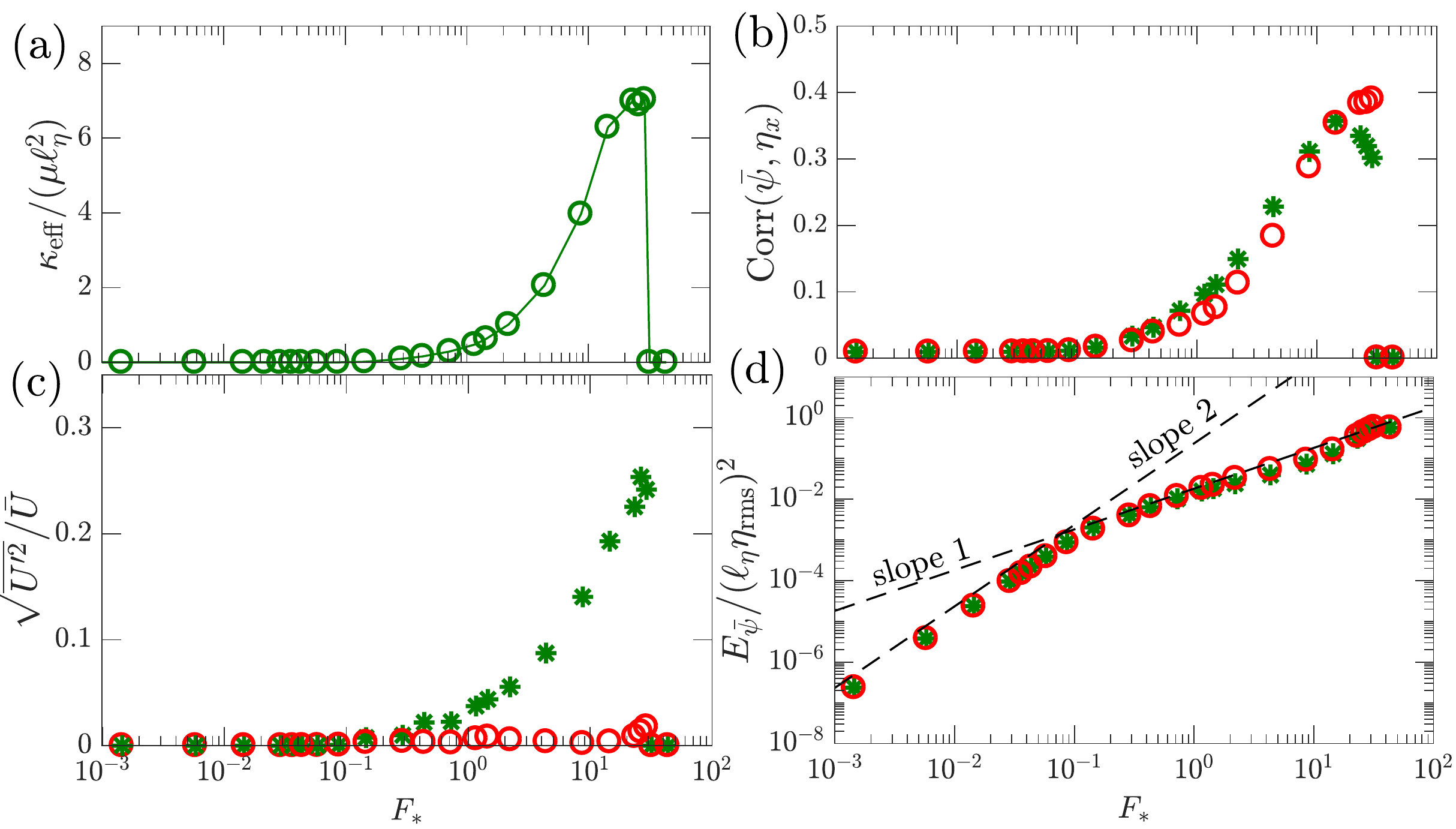}
\vspace{-0.5em}\caption{(a) The effective PV diffusivity, $\keff$, diagnosed from~\eqref{eq:kappa_energ_mean} for the series of solutions with $\bs=1.38$. (b) The correlation $\corr(\bar\psi,\eta_x)$ for this series of solutions (asterisks~$*$) and for the solutions with parameterized transient eddies (circles~$\circ$). (c) Same as panel (b) but showing the strength of the transient eddies using the index \eqref{usindex}. (c) Same as panel (b), but showing the energy of the standing eddies $\bar \psi$. Also shown are the power laws $\Fs^1$ and $\Fs^2$ as dashed black lines.}
\label{fig:QL_b20_keff}
\end{figure}

Figure~\ref{fig:QL_b20_nueff_1p} compares the numerical solutions of~\eqref{eq:U_t} and~\eqref{eq:z_NL} with the prediction of the \quasilinear{} approximation (asterisks $*$ versus the solid QL curve) for the case with $\bs=1.38$. The \quasilinear{} approximation has  a stronger  large-scale flow than that of the full system in~\eqref{eq:U_t} and~\eqref{eq:z_NL}.  Moreover the full system is more impressively eddy saturated than the QL approximation. There are at least two causes for these failures of the \quasilinear{} approximation: \textit{(i)} QL assumes steady flow and has no way of incorporating the effect of transient eddies on the time-mean flow and \textit{(ii)} QL neglects the term $\J(\bar\psi,\bar q)$.

We address these points by following \citet{Rhines-Young-1982} and approximating the effect of the transient eddies as PV diffusion:
\beq
 \bnabla\bcdot\bE \approx -\keff\lap \bar{q}\per\label{eq:PVdiff}
\eeq
In the discussion surrounding~\eqref{eq:keff_app}, we determine $\keff$ using the time-mean eddy energy power integral \eqref{eq_Epsid}.  According to this diagnosis, the PV diffusivity is
\begin{align}
	\keff = \mu \overline{\la |\grad\psi'|^2\ra}\Big/\laa \bar{\zeta}\,\bar{q}\raa\per
	\label{eq:keff_simple}
\end{align}
 Figure~\ref{fig:QL_b20_keff}(a) shows $\keff$ in \eqref{eq:keff_simple} for the solution suite with $\bs = 1.38$.

 \citet{Abernathey-Cessi-2014}, in their study of baroclinic equilibration in a  channel with topography, developed a two-layer QG model that incorporated the role of standing eddies in determining the transport. \citet{Abernathey-Cessi-2014} also used an effective PV diffusion to parametrize transient eddies. However, they specified $\keff$, rather than determining it diagnostically from the energy power integral as in~\eqref{eq:keff_simple}.


With $\keff$ in hand, we can revisit the \quasilinear{} theory and ask for its prediction when the term $\keff\nabla^2 q$ is added on the right hand side of~\eqref{eq:z_QL}. This way we include the effect of the transients on the time-mean flow but do not include the effect of the term $\J(\bar\psi,\bar q)$. The \quasilinear{} prediction is only slightly improved --- see the dash-dotted curve in figure~\ref{fig:QL_b20_nueff_1p}.

To include also the effect of the term $\J(\bar\psi,\bar q)$ we obtain solutions of~\eqref{eq:U_t} and~\eqref{eq:z_NL} with added PV diffusion in~\eqref{eq:z_NL} with $\keff$ as in figure~\ref{fig:QL_b20_keff}(a). We find that the strength of the transient eddies is dramatically reduced: see figure~\ref{fig:QL_b20_keff}(c). Thus the approximation~\eqref{eq:PVdiff} with the PV diffusivity supplied by~\eqref{eq:keff_simple} is self-consistent in the sense that we do not both resolve and parameterize transient eddies. Moreover, the large-scale flow $\bar U$ with parameterized transient eddies is in much closer agreement with $\bar U$ from the solutions with transient eddies --- see figure~\ref{fig:QL_b20_nueff_1p}. This striking quantitative agreement as we vary $\Fs$ shows that at least in the case with $\bs =1.38$ the transient eddies act as PV diffusion on the time-mean flow.

Thus we conclude, that in addition to $\beta$, the main physical mechanisms operating in the eddy saturation regime are PV diffusion via the transient eddies and the mean advection of mean PV i.e., the term $\J(\bar \psi, \bar q)$.

There are two  remarkable aspects of this success. First, it is important to use $\keff \lap (\bar \zeta + \h)$ in~\eqref{eq:PVdiff}; if one uses only $\keff\lap \bar \zeta$ then the agreement in figure~\ref{fig:QL_b20_nueff_1p} is degraded. Second, PV diffusion does not decrease the amplitude of the standing eddies: see figure~\ref{fig:QL_b20_keff}(d). Furthermore, PV diffusion quantitatively captures  $\corr(\bar \psi,\h_x)$: see figure~\ref{fig:QL_b20_keff}(b).

Unfortunately, the success of the PV diffusion parameterization  does not extend to cases with closed geostrophic contours (small $\bs$), such as  $\bs=0.10$. For small $\bs$ the flow is strongly affected by the detailed structure of the topography. The solution described in section~\ref{sec:example_closed} shows that flow is channeled into a few streams and thus a parametrization that does account the actual structure of the topography is, probably, doomed to fail. In fact, for $\bs=0.1$ the $\keff$ diagnosed according to~\eqref{eq:keff_simple} is negative because $\la\bar\zeta \bar q\ra<0$.

In conclusion, the PV diffusion approximation~\eqref{eq:PVdiff} gives good quantitative results provided that  the flow does not crucially dependent on the structure of the topography itself i.e., for large $\bs$ so that the geometry is dominated by open geostrophic contours. In the context of baroclinic models, eddy saturation is not captured by standard  parameterizations of transient  baroclinic eddies \citep{Hallberg-Gnanadesikan-2001}. Only very recently have \citet{Mak-etal-2017} proposed a parameterization of baroclinic turbulence that successfully  produces baroclinic eddy saturation. Thus the success of \eqref{eq:PVdiff} in the barotropic context, even though it depends on diagnosis of $\keff$ via \eqref{eq:keff_simple}, is significant.

\subsection{Drag crisis\label{drag_cr}}

In this section we provide some further insight into the drag crisis. We argue that the requirement of enstrophy balance among the flow components leads to a transition from the lower to the upper branch as wind stress forcing increases. We make this argument by constructing lower bounds on the large-scale flow $\bar U$ based on energy and enstrophy power integrals.

We consider a ``test streamfunction'' that is efficient at producing form stress:
\beq
	\psit = \alpha\h_x\com\label{eq:test_psi}
\eeq
with $\alpha$ a positive constant to be determined by satisfying  either the energy or the enstrophy power integrals from  appendix \ref{app:balances}.  A maximum form stress corresponds to a minimum large-scale flow $\Umin$, which in turn can be determined by substituting~\eqref{eq:test_psi} into the time-mean large-scale flow  equation~\eqref{eq:U_NL_static}:
\beq
	\mu \Umin= F - \alpha\la\h_x^2\ra\per
\label{eq:cri3}
\eeq

We can determine $\alpha$ so that the eddy energy power integral~\eqref{eq_Epsid}+\eqref{eq_Epsim} is satisfied:
\beq
	0=\Umin \,\alpha \la\h_x^2\ra - \mu\alpha^2 \la|\grad\h_x|^2\ra - \nu\alpha^2\la(\lap  \h_x)^2\ra\per
\label{eq:cri1}
\eeq
The averages above are evaluated using properties of monoscale topography, e.g. $\la(\lap\h_x)^2\ra = \half\etarms^2 \leta^{-6}$. Solving~\eqref{eq:cri3} and~\eqref{eq:cri1} for $\alpha$ and $\Umin$ we obtain a lower bound on the large-scale flow based on the energy constraint,
\begin{align}
	\bar{U} \ge \UminE \defn \frac{F}{\mu} \[ 1+\frac{\etarms^2}{2(\mu+\nu/\leta^2)^2} \]^{-1}\per\label{Ebnd}
\end{align}
Alternatively, one can determine $\alpha$ and $\Umin$ by satisfying  the eddy enstrophy power integral~\eqref{eq_Qpsim}+\eqref{eq_Qpsid}.  This leads to a second bound,
\begin{align}
	\bar{U}\ge \UminQ\defn  \frac{F}{\mu}  \[ 1-\frac{\beta\etarms^2\leta^2}{2(\mu+\nu/\leta^2) F} 	\] \per\label{Qbnd}
\end{align}
Thus
\beq
	U \ge \max\left( \UminE, \UminQ\right)\per\label{bounders}
\eeq
The test function in~\eqref{eq:test_psi} does not closely resemble the realized flow so the bound above is not tight. Nonetheless it does capture some qualitative properties of the turbulent solutions.

(Using a more elaborate test function with two parameters one can satisfy both the energy and enstrophy power integrals simultaneously  and obtain a single bound.  However, the calculation is much longer and the result is not much better than the relatively simple~\eqref{bounders}.)

\begin{figure}
	\centering
	\includegraphics[width = .7\textwidth]{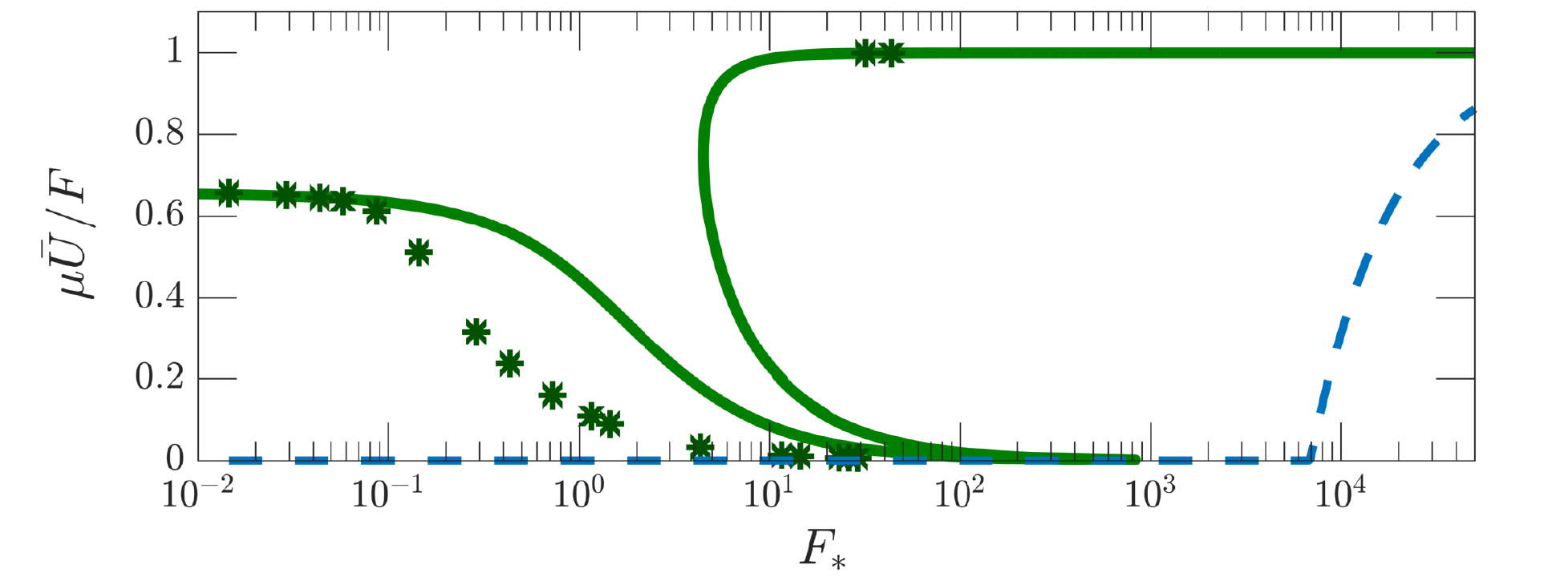}
	\vspace{-0.5em}\caption{The numerical solutions of~\eqref{eq:U_t} and~\eqref{eq:z_NL} (asterisks $*$) and the \quasilinear{} prediction~\eqref{eq:sigma_QLf} (solid curve). The dashed line shows the lower bound~\eqref{bounders}.}\label{fig:bounds}
\end{figure}

The lower bound~\eqref{bounders} is shown in figure~\ref{fig:bounds} for the case with $\bs=1.38$ together with the numerical solution of the full nonlinear equations~\eqref{eq:U_t} and~\eqref{eq:z_NL} and the \quasilinear{} prediction. Although it cannot be clearly seen, the energy bound $\UminE$ does not allow $\bar{U}$ to vanish completely, e.g. for the $\mus=10^{-2}$ we have that $\mu\UminE/F= 2\times10^{-4}$. On the other hand, the dominance of the enstrophy bound  $\UminQ$ at high forcing  explains the  occurrence of the drag crisis: the enstrophy power integral  \emph{requires} that the large-scale flow transitions from the lower to the upper branch as $F_*$ is increased beyond a certain value.

These bounds provide a qualitative explanation  for the existence the drag crisis.  The critical forcing  predicted by the enstrophy bound dominance overestimates the actual value of the drag crisis. For example, for the case with $\bs=1.38$ shown in figure~\ref{fig:bounds},  $\UminQ$ becomes the lower bound at a value of  $\Fs$ that is about 240 times larger than the actual drag crisis point. We have no reason to expect these bounds to be tight:  they do not depend on the actual structure of topography itself but only on gross statistical properties, e.g. $\etarms$, $\leta$, $\Leta$. For example, a sinusoidal topography in the form of
\beq
  \h = \sqrt{2} \etarms\cos{(x/\leta)}\label{eq:PVcosmx}
\eeq
has identical statistical properties as the random monoscale topography used in this paper and, therefore, imposes the same bounds. But with the  topography in~\eqref{eq:PVcosmx} there is a laminar solution with $\psi_y=0$ and, as a result, also $\J(\psi,q)=0$. In this case the \quasilinear{} solution~\eqref{eq:psi_QLf} is an \emph{exact} solution of the full nonlinear equations~\eqref{eq:U_t} and~\eqref{eq:z_NL} and the bound~\eqref{bounders} is tight.

\section{Discussion and conclusion \label{conclusion}}

The main new results in this work are illustrated by  the two limiting cases described in sections \ref{sec:example_closed} and \ref{sec:example_open}.  The   case  in section \ref{sec:example_closed}, with $\bs=0.1$, is realistic in  that ballpark estimates indicate that topographic  PV will overpower $\beta y$ to produce closed geostrophic contours almost everywhere. The topographically blocked flow then consists of close-packed stagnant ``dead zones'' separated by narrow jets. Dead zones are particularly   notable in the steady  solution shown in  figure \ref{fig:Nu_mono}. But they are also clear in the unsteady solution   of figure \ref{fig:eg_closed_B}. There is no eddy saturation in this topographically blocked regime: the large-scale flow $\bar U$ increases roughly linearly with $F$ till the drag-crisis jump to the upper branch. Because of the topographic partitioning into dead zones,  the large-scale time-mean flow $\bar U$ is an unoccupied mean: in most of the domain there are  weak recirculating eddies.

The complementary limit, illustrated by the case  in section \ref{sec:example_open} with $\bs=1.38$, is when all of the geostrophic contours are open. In this limit  we find that: \textit{(i)}  the large-scale flow $\bar U$ is insensitive to changes in  $F$;  \textit{(ii)}  the kinetic energy of the transient eddies increases linearly with $F$: see figure \ref{fig:QL_b20_keff}(d). Both \textit{(i)}  and \textit{(ii)}  are defining  symptoms of the eddy-saturation phenomenon documented in eddy resolving Southern-Ocean models. \citet{Constantinou-2017} identifies a third symptom common to  barotropic and baroclinic eddy saturation: increasing the Ekman drag coefficient $\mu$ increases the  large-scale mean flow \citep{Marshall-etal-2017}. This somewhat counterintuitive dependence on $\mu$  can be rationalized by arguing that the main effect of increasing Ekman drag is to damp the transient eddies responsible for producing $\keff$ in   \eqref{eq:keff_simple}. Section \ref{sec:saturation} shows that decreasing the strength of the transient eddies, and their associated effective PV diffusivity,  also decreases the topographic form stress and thus   increases the transport.

The two limiting cases   described above are not quirks of  the monoscale topography in figure \ref{fig:topo}: using a multiscale topography with a $k^{-2}$ power spectral density, we find similar qualitative behaviors  (not shown here), including eddy saturation in the limit of section \ref{sec:example_open}. Moreover, the main controlling factor for  eddy saturation in this barotropic model is whether the geostrophic contours are open or closed --- the numerical value of $\bs$ is important only in so far as $\bs$ determines whether the geostrophic contours are open or closed. For example, the ``unidirectional'' topography in \eqref{eq:PVcosmx} always has open geostrophic contours;  using this sinusoidal topography \citet{Constantinou-2017} shows that there is barotropic   eddy saturation with $\bs$ as low as $0.1$.  Thus it is the structure of the geostophic contours, rather than the numerical value of $\bs$, that  is decisive as far as barotropic eddy saturation is concerned.

The  explanation of baroclinic eddy saturation, starting with \citet{Straub-1993},  is that isopycnal slope  has a hard upper limit  set by the marginal condition for baroclinic instability. As the strength of the wind is increased from small values, the isopycnal slope initially increases and so does the associated  ``thermal-wind transport''. (The thermal-wind transport is diagnosed from the density field   by integrating the thermal-wind relation upwards from a level of no motion at the bottom.) However, once the isopycnal slope reaches    the marginal condition for baroclinic instability, further increases in slope, and in thermal-wind transport, are no longer possible. At the margin of baroclinic instability,   the unstable  flow can easily make more eddies to counteract further wind-driven steepening of  isopycnal slope. This is the standard explanation of  baroclinic eddy saturation in which the transport (approximated by the thermal-wind transport) is unchanging, while the strength of the transient eddies increases linearly with wind stress.

Direct comparison of the barotropic model  with baroclinic Southern-Ocean models, and with the Southern  Ocean itself, is difficult and probably not worthwhile except for gross parameter estimation as in table \ref{tab:SOvalues}.  However several qualitative points should be mentioned. Most strikingly, we find that eddy saturation occurs without baroclinic instability and without thermal-wind transport. This finding challenges the standard explanation of eddy saturation in terms of the marginal condition for baroclinic instability. Nonetheless, the onset of transient barotropic eddies, shown in figure \ref{fig:dUm_Um}(b), is also the onset of barotropic  eddy saturation. This barotropic-topographic instability is the source of the transient eddies that produce $\keff$ in \eqref{eq:keff_simple}.  Thus one can speculate that barotropic eddy saturation  also involves a flow remaining close to a marginal stability condition. Substantiating this claim, and clarifying the connection  between barotropic and baroclinic eddy saturation, requires better characterization  the barotropic-topographic instability and also of the effect of small-scale topography on baroclinic instability. The latter point is fundamental: in a baroclinic flow topographically blocked  geostrophic contours in the deep layers co-exist with open contours in  shallower layers. The marginal condition for baroclinic instability in this circumstance is not well understood. The issue is further confused because transient eddies generated by barotropic-topographic  instability have the same length scale as the topography, which can be close to the deformation length scale of baroclinic eddies.

%
%



\defcitealias{Ward-Hogg-2011}{Ward \& Hogg}

Further evidence for the importance of barotropic processes in establishing eddy saturation is  provided by several Southern-Ocean type models. \citet{Abernathey-Cessi-2014} showed that an isolated ridge results in localized baroclinic instability  over the ridge and a downstream barotropic standing wave. Relative to the flat-bottom case, transient eddies are weak in most of the domain and the thermocline is shallow with small slopes; see  \citet{Thompson-NaveiraGarabato-2014} for further discussion of the role of barotropic standing waves in setting the momentum balance and transport in the Southern  Ocean. In a study of channel spin-up, \citet{Ward-Hogg-2011} showed that a suddenly imposed  wind stress is balanced by topographic form stress within  two or three weeks.  This fast balance is achieved by barotropic pressure gradients associated with sea-surface height. Interior equilibration, involving transmission of momentum by  interfacial form stresses and baroclinic instability \citep{Johnson-Bryden-1989}, takes about 10 years to establish. Wind stress on the Southern Ocean   is never steady and  thus fast barotropic eddy saturation  may be as important as slow baroclinic eddy saturation.

\vspace{1em}

The authors acknowledge fruitful discussions with N.\;A.~Bakas, P.~Cessi, T.\;D.~Drivas, B.\;F.~Farrell, G.\;R.~Flierl, A.\;McC. Hogg, P.\;J.~Ioannou and A.\;F.~Thompson. Comments from three anonymous reviewers greatly improved the paper. NCC~was supported by the NOAA Climate \& Global Change Postdoctoral Fellowship Program, administered by UCAR's Cooperative Programs for the Advancement of Earth System Sciences. WRY was supported by the National Science Foundation under OCE-1357047. The \MATLAB code used in this paper is available at the github repository: \url{https://github.com/navidcy/QG_ACC_1layer}.

\appendix

\section{Energy and enstrophy power integrals and balances\label{app:balances}}

In this appendix we derive energy and enstrophy power integrals as well as the time-averaged energy and enstrophy balances for each of the four flow components: $\bar U$, $U'$, $\bar \psi$ and $\psi'$.


The energy and enstrophy of the flow are defined in~\eqref{eq:E_Q_defs}. From~\eqref{eq:U_t} and~\eqref{eq:z_NL} we find that:
\begin{subequations}
\begin{align}
\frac{\dd E}{\dd t} &= \wind U - \mu U^2  -   \laa\mu |\grad \psi |^2  + \nu\zeta^2 \raa \com \label{eq:dEdt}\end{align}

\begin{align}
\frac{\dd Q}{\dd t} &= \wind\beta - \laa \eta \D \zeta\raa -\mu \beta U - \laa \mu \zeta^2 + \nu |\grad \zeta|^2 \raa \per \label{eq:dQdt}
\end{align}\label{eq:dEdQdt}\end{subequations}
The rate of working by the wind stress, $\wind U$, appears on the right of~\eqref{eq:dEdt}: because $F$ is constant  the energy injection  varies directly  with the large-scale mean flow $U(t)$. On the other hand, the main enstrophy injection rate  on the right of~\eqref{eq:dQdt} is fixed and equal to $\wind\beta$.
The subsidiary enstrophy source $\la\eta \D \zeta\ra$ becomes important if $\beta$ is small relative to the gradients of the topographic PV; in the special case $\beta=0$, $\la \eta \D \zeta\ra$ is the only enstrophy source.


 Following~\eqref{tStand}, we represent all flow fields as a time-mean plus a transient; note that $\bar q = \bar \zeta + \h$ and $q'=\zeta'$. Equations~\eqref{eq:U_t} and~\eqref{eq:z_NL} decompose into:\begin{subequations}
\begin{gather}
\J(\bar{\psi}-\bar U y,\bar{q}+\beta y) + \bnabla \bcdot \bE + \D \bar\zeta = 0\com\label{eq:B3}\\
q'_t   + \J(\psi'-U'y,\bar{q}+\beta y)  + \J(\bar{\psi}-\bar{U}y,q') + \bnabla \bcdot \left( \bE'' - \bE\right) + \D \zeta' = 0\label{eq:B4}\com\\
	 \wind -\mu \bar{U}  - \fsb =0\com  \label{eq:B1}\\
	 U'_t =-\mu U'  - \la\psi'\h_x\ra \label{eq:B2}\com
 \end{gather}\label{eqsBs}\end{subequations}
where the eddy PV fluxes are $\bE'' \defn \bU'q'$ and  $\bE \defn \overline{\bU'q'}$.

\subsection{Energy and enstrophy balances\label{sec:app_energ}}

Following the definitions in~\eqref{eq:E_Q_defs}, the energy of each flow component is:
\beq
E_{\bar U} = \half \bar{U}^2\ \ ,\ \ E_{U'} = \half {U'}^2\ \ ,\ \ E_{\bar \psi} = \half \la|\grad\bar\psi|^2\ra\ \ \text{and}\ \  E_{\psi'} = \half \la|\grad \psi'|^2\ra\per
\eeq
Thus the total energy of the large-scale flow and of the eddies is:
\begin{gather}
E_{U} = E_{\bar U} + E_{ U'} + \bar U\,U'\andd
E_{\psi} = E_{\bar \psi} + E_{\psi'} + \la\grad\bar\psi\bcdot\grad\psi'\ra\per
\end{gather}
The cross-terms above are removed by time-averaging.

The time-mean energy balances for each flow component are obtained by manipulations of~\eqref{eqsBs} as follows:
\begin{subequations}
	\begin{align}
		E_{\bar\psi}&:\qquad  \la -\bar\psi \times \text{\eqref{eq:B3}} \ra &\Rightarrow&\quad   0=\bar{U}\fsb + \la\bar{\psi}\grad\bcdot\bE\ra + \la\bar\psi\D\bar\zeta\ra \com\label{eq_Epsim}\\
		\overline{E_{\psi'}}&:\qquad \overline{\la -\psi' \times \text{\eqref{eq:B4}} \ra} &\Rightarrow&\quad   0=\overline{U'\la\psi'\h_x\ra} -\la\bar{\psi}\grad\bcdot\bE\ra + \overline{\la\psi'\D\zeta'\ra} \label{eq_Epsid}\com\\
		E_{\bar{U}}&:\qquad \bar{U} \times \text{\eqref{eq:B1}} &\Rightarrow&\quad   0 =F\bar{U}-\mu\bar{U}^2- \bar{U}\fsb \label{eq_EUm}\com\\
		\overline{E_{U'}}&:\qquad\overline{U'\times \text{\eqref{eq:B2}}} &\Rightarrow&\quad  0 =  -\mu \overline{U'^2}- \overline{U'\la\psi'\h_x\ra} \label{eq_EUd}\per
	\end{align}\label{eqs:Ebal}\end{subequations}
Summing~\eqref{eq_EUd} and~\eqref{eq_EUm} we obtain the energy power integral for the total (standing plus transient) large-scale flow. Summing~\eqref{eq_Epsid} and~\eqref{eq_Epsim} the conversion term $\la\bar{\psi}\grad\bcdot\bE\ra$ cancels and we obtain the energy power integral~\eqref{eq:U_fs} for the total (standing plus transient) eddy field.

\begin{figure}
\centering
\includegraphics[width = .7\textwidth]{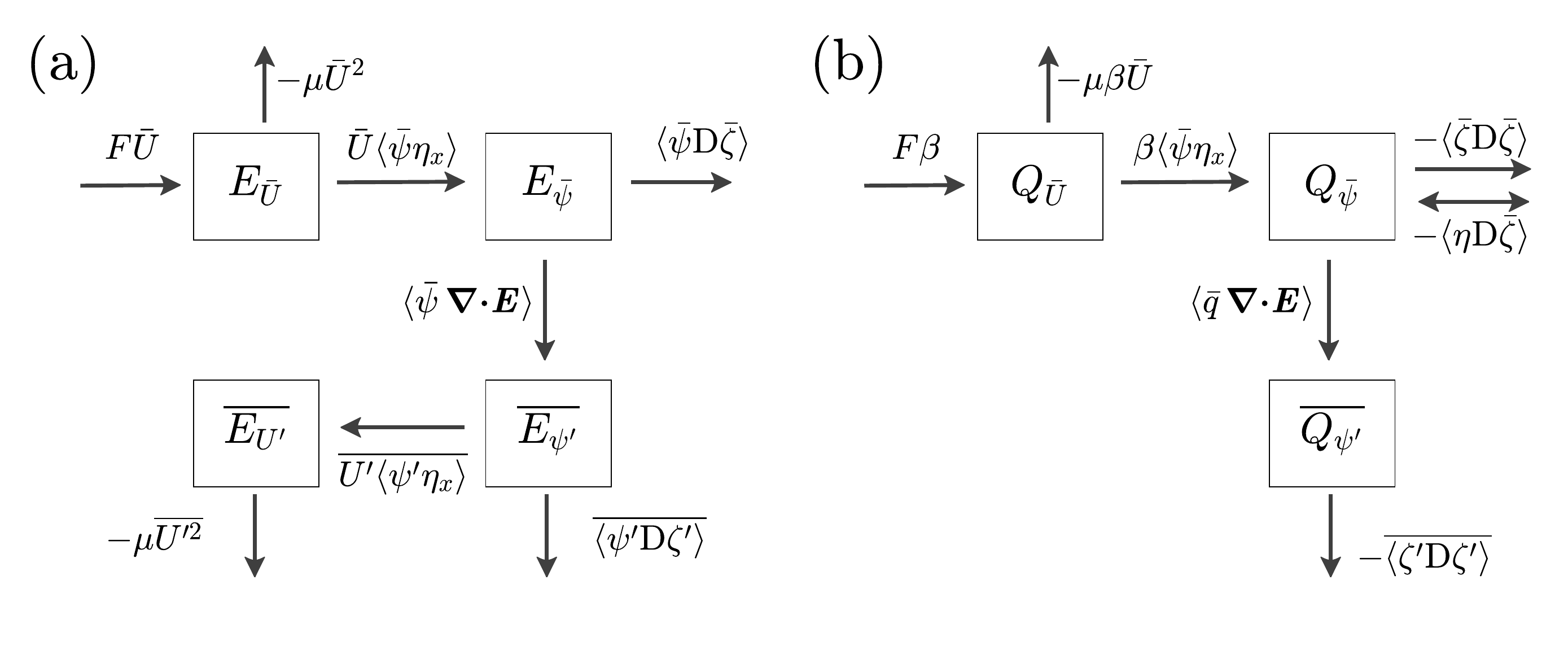}
\vspace{-2em}
\caption{The energy and enstrophy transfers between  the four flow components: the time-mean large-scale flow $\bar{U}$, the standing eddies $\bar{\psi}$, and  the corresponding transient components $U'$ and $\psi'$.}
\label{fig:E_Q_diagram}
\end{figure}

The time-mean of the energy integral in~\eqref{eq:dEdt} is the sum of equations~\eqref{eqs:Ebal}. Note that from~\eqref{eq_EUd} we have that $\overline{U'\la\psi'\h_x\ra}<0$ and thus from~\eqref{eq_Epsid} we infer that $\la\bar{\psi}\grad\bcdot\bE\ra<0$. The energy balances~\eqref{eqs:Ebal} are summarized in figure~\ref{fig:E_Q_diagram}(a).

In section~\ref{sec:saturation} we approximate $\grad\bcdot\bE$ as PV diffusion~\eqref{eq:PVdiff}. The effective PV diffusivity $\keff$ can be diagnosed by  requiring  that  the time-mean eddy energy balance~\eqref{eq_Epsim} and the transient eddy flow energy balance~\eqref{eq_Epsid} are satisfied. according to this requirement gives:\begin{subequations}
\begin{align}
\keff &=\(\bar U\la\bar\psi\eta_x\ra + \la\bar\psi\D\bar\zeta\ra\)\Big/\laa \bar{\zeta}\,\bar{q}\raa \label{eq:kappa_energ_mean1} \\
&=-\(\vphantom{\bar U}\right. \overline{\la \psi'\D\zeta'\ra} + \overline{U'\la\psi'\eta_x\ra}\left.\vphantom{\bar U}\)\Big/\laa \bar{\zeta}\,\bar{q}\raa\label{eq:kappa_energ_mean}\per
\end{align}\label{eq:keff_app}\end{subequations}
In \eqref{eq:kappa_energ_mean1} the terms $\bar U\la\bar\psi\eta_x\ra$ and $\la\bar\psi\D\bar\zeta\ra$ are of opposite sign; the magnitude of the former is generally much larger than that of the latter. In \eqref{eq:kappa_energ_mean}, the term $\overline{U'\la\psi'\eta_x\ra}$ is negligible compared to $\overline{\la \psi'\D\zeta'\ra}$. Neglecting the small term, and using $\D \zeta'=\mu \zeta'$, we  simplify  \eqref{eq:kappa_energ_mean} to obtain the expression for $\keff$ in~\eqref{eq:keff_simple}.

The enstrophy of each flow component is:
\beq
Q_{\bar U} = \beta \bar{U}\ \ ,\ \ Q_{U'} = \beta U'\ \ ,\ \ Q_{\bar \psi} = \half \la \bar{q}^2\ra\ \ \text{and}\ \ Q_{\psi'} = \half \la q'^2\ra\com
\eeq
The transient large-scale flow has by definition $\overline{Q_{U'}}=0$. The enstrophy power integrals follow by manipulations similar to those in~\eqref{eqs:Ebal}:\begin{subequations}
\begin{align}
	 Q_{\bar{\psi}}&:\qquad  \la \bar q \times \text{\eqref{eq:B3}} \ra &\Rightarrow&\quad  0 = \beta \la\bar\psi\h_x \ra - \la\bar q\,\grad\bcdot\bE \ra - \la\bar \z \D \bar\zeta\ra -\la\h \D \bar\zeta\ra  \com\label{eq_Qpsim}\\
	\overline{Q_{\psi'}}&:\qquad  \overline{\la q' \times \text{\eqref{eq:B4}}} \ra &\Rightarrow&\quad  0 =  \laa \bar{q} \,\grad \bcdot \bE\raa  - \overline{ \laa\zeta'\D\zeta'\raa}\com\label{eq_Qpsid}\\
	Q_{\bar{U}}&:\qquad  \la \beta \times \text{\eqref{eq:B1}} \ra &\Rightarrow&\quad 0= F\beta -\mu \beta\bar{U}-\beta\fsb\per\label{eq_QUm}
\end{align}\label{eqs:Qbal}\end{subequations}
The time-mean of the enstrophy integral in~\eqref{eq:dQdt} is the sum of equations~\eqref{eqs:Qbal}. Equation~\eqref{eq_Qpsid} implies that $\la\bar{q}\,\grad\bcdot\bE\ra>0$;  the term $\la\h\D\bar{\zeta}\ra$ in~\eqref{eq_Qpsim} can have either sign. The enstrophy power integrals~\eqref{eqs:Qbal}  are summarized  in figure~\ref{fig:E_Q_diagram}(b).

\def\psitot{\Psi}
\def\qtot{Q}

\def\psitot{\psi_{\rm tot}}
\def\qtot{q_{\rm tot}}

\def\psitot{\Psi}
\def\qtot{\bar q + \beta y}

\section{Form stress for isotropic topography}\label{app:fs_mono}

For the case of isotropic topography analytic progress follows to the \quasilinear{} expression for the form stress by converting the sum over $\bk$ in~\eqref{eq:sigma_QLf} into an integral:
\beq
	\la\psi\eta_x\ra = U \int \frac{ \mu k_x^2|\bk|^2 |\hat{\h}(\bk)|^2}{\mu^2 |\bk|^4 + k_x^2 (\beta-|\bk|^2U)^2}\, \dd^2\bk \com\label{eq:fs_QL_int}
\eeq
where $\hat{\h}(\bk) \defn \int \h(\bx)\ee^{-\ii\bk\bcdot\bx}\,\dd^2\bx$. Now assume that the topography is isotropic, i.e. its power spectral density $S(k)$ is only a function of the total wavenumber $k=|\bk|$ and γιωεν as $S(k)= 2\upi k |\hat{\h}(\bk)|^2$, so that $\etarms^2=\int S(k)\,\dd k$. In this case   we further simplify the integral~\eqref{eq:fs_QL_int} using polar coordinates $(k_x,k_y)=k (\cos\theta\, ,\sin\theta)$:
\beq
	\la\psi\eta_x\ra = \frac{U}{2\upi\mu} \int_{0}^\infty\!\!\!  \,  S(k)\oint\frac{ \cos^2\theta }{1  + \xi \cos^2\theta} \, \dd\theta\,\dd k\com
\eeq
where $\xi \defn \left(\beta -k^2 U\right)^2/(\mu k)^2>0$.
The $\theta$-integral above is evaluated analytically so  that
\beq
	\la\psi\eta_x\ra =  \mu U \int_{0}^\infty\!\!\frac{k^2 S(k)\,\dd k}{\mu^2k^2 +\left(\beta -k^2 U\right)^2+\mu k\sqrt{\mu^2k^2 + \left(\beta -k^2 U\right)^2}}\per\label{eq:A5}
\eeq

For the special case of idealized monoscale topography: $S(k) = \etarms^2\,\delta \left(k-\leta^{-1}\right)$, the $k$-integral in~\eqref{eq:A5} can be evaluated in closed form. In that case~\eqref{eq:A5} reduces to~\eqref{eq:sigma_QLf_mono}.

\end{document}